\newcommand{\version}{September 14, 2011}
\documentclass[11pt,a4paper,english,pdftex]{article}

\usepackage[bindingoffset=0.3cm,textheight=22.5cm,hdivide={2.7cm,*,2.7cm}, vdivide={*,22cm,*}]{geometry}
\usepackage[pdftex,bookmarksnumbered=true,breaklinks=true]{hyperref}

\usepackage{amsmath,amsfonts,amssymb,slashed,mathrsfs}

\usepackage[numbers,square,comma,sort&compress]{natbib}

\IfFileExists{dsfont.sty}
	{\usepackage{dsfont}
         \newcommand{\id}{\mathds{1}}}
	{\typeout{Package dsfont.sty was not found, using alternative macros.}
         \let\mathds=\mathbb
         \newcommand{\id}{\mbox{1 \kern-.59em {\rm l}}}}
\let\one=\id

\makeatletter

         \@addtoreset{equation}{section}
\makeatother

\newcommand{\nocontentsline}[3]{}
\newcommand{\tocless}[3]{\bgroup\let\addcontentsline=\nocontentsline#1{#2}#3\egroup}

\newcommand{\Appendix}[1]{
  \refstepcounter{section}
  \section*{Appendix \thesection:\hspace*{1.5ex} #1}
  \addcontentsline{toc}{section}{Appendix \thesection}
}
\newcommand{\SubAppendix}[2]{\tocless\subsection{#1}{\label{#2}}}

\newcommand{\qed}{\nobreak \ifvmode \relax \else
      \ifdim\lastskip<1.5em \hskip-\lastskip
      \hskip1.5em plus0em minus0.5em \fi \nobreak
      \vrule height0.75em width0.5em depth0.25em\fi}

\newcommand{\be}{\begin{equation}}
\newcommand{\ee}{\end{equation}}
\newcommand{\eq}[1]{(\ref{#1})}

\def\nn{\nonumber}
\def\bea{\begin{eqnarray}}
\def\eea{\end{eqnarray}}
\def\obar{\overline}

\def\beqa{\begin{eqnarray}} 
\def\eeqa{\end{eqnarray}} 
\def\beq{\begin{equation}} 
\def\eeq{\end{equation}}

\def\Tr{{\rm Tr}}

\def\a{\alpha}          
\def\b{\beta}           

\def\l{\lambda} \def\L{\Lambda} 

\def\r{\rho}

\def\cA{{\cal A}}  \def\cC{{\cal C}}
  \def\cF{{\cal F}}
\def\cG{{\cal G}} \def\cH{{\cal H}} 
 \def\cK{{\cal K}} \def\cL{{\cal L}}
\def\cM{{\cal M}} \def\cN{{\cal N}} \def\cO{{\cal O}}

\def\mg{\mathfrak{g}}

\def\algA{\mathscr{A}}

\newcommand{\bG}{\bar{G}}

\newcommand{\bc}{\bar c}

\newcommand{\R}{\mathds{R}}
\newcommand{\C}{\mathds{C}}

 \newcommand{\msu}{\mathfrak{s}\mathfrak{u}}
 \newcommand{\mso}{\mathfrak{s}\mathfrak{o}}
\newcommand{\mmu}{\mathfrak{u}}
\newcommand{\mSU}{SU}
\newcommand{\mU}{U}

\def\bit{\begin{itemize}}
\def\eit{\end{itemize}}

\def\({\left(}
\def\){\right)}
\def\diag{\mbox{diag}}

\def\Mat{{\rm Mat}}

\def\pa{\partial} \def\del{\partial}

\newcommand{\tr}{\mbox{tr}}

\def\bcomment#1{}

\def\LNC{\L_{\rm NC}}
\newcommand{\bLNC}{\bar\L_{\rm NC}}

\newcommand{\Sin}[1]{\sin\!\left(\! \frac{ #1 }{2} \!\right)}
\newcommand{\Cos}[1]{\cos\!\left(\! \frac{ #1 }{2} \!\right)}

\newcommand{\Intt}[1]{\int\!\!\frac{d^4 #1}{(2\pi\LNC^2)^2}}
\newcommand{\dott}[2]{#1 \!\cdot\! #2}
\newcommand{\Dott}[1]{#1 \!\cdot\! #1}

\newcommand{\nc}{non-com\-mu\-ta\-tive}

\newcommand{\eqnref}[1]{Eqn.~(\ref{#1})}		
\newcommand{\secref}[1]{Section~\ref{#1}}		
\newcommand{\appref}[1]{Appendix~\ref{#1}}		
\newcommand{\inv}[1]{\frac{1}{#1}}				
\newcommand{\tinv}[1]{\tfrac{1}{#1}}
\newcommand{\co}[2]{[#1,#2]}						
\newcommand{\var}[2]{\frac{\d #1}{\d #2}}				
\newcommand{\intx}{\int\! d^4x}						
	
\renewcommand{\Re}{\textrm{Re}}

\newcommand{\effM}{\mathfrak{m}}

\renewcommand{\ln}{\log}				

\renewcommand{\a}{\alpha}
\renewcommand{\b}{\beta}
\newcommand{\g}{\gamma}
\renewcommand{\d}{\delta}

\renewcommand{\th}{\theta}
\newcommand{\thb}{\bar\theta}

\renewcommand{\l}{\lambda}
\newcommand{\m}{\mu}
\newcommand{\n}{\nu}

\renewcommand{\r}{\rho}
\newcommand{\s}{\sigma}

\newcommand{\vG}{\varGamma}

\newcommand{\Th}{\Theta}

\renewcommand{\L}{\Lambda}

\renewcommand{\Xi}{\Xi}

\title{\begin{flushright}
       \small{UWThPh-2011-29}
       \end{flushright}
\vspace{3em}
\bf
On the 1-loop effective action for the IKKT model \\[1ex] and non-commutative branes}
\author{Daniel N. Blaschke\footnote{daniel.blaschke@univie.ac.at}~, Harold Steinacker\footnote{harold.steinacker@univie.ac.at}}

\date{\version}
\begin{document}

\maketitle

\begin{center}
\textit{University of Vienna, Faculty of Physics\\
Boltzmanngasse 5, A-1090 Vienna (Austria)}
\vspace{0.5cm}
\end{center}%
\begin{abstract}

We study the one-loop effective action of the IKKT or IIB model on a 4-dimensional non-commutative brane background.
The trace--$U(1)$ sector is governed by non-commu\-ta\-tivity, and leads 
--- assuming no SUSY breaking --- to a higher-derivative effective action. 
In contrast, the non-Abelian sector at low energies reduces to $SU(N)$ $\cN=4$ Super-Yang-Mills on the brane,
with a global $SO(9,1)$ symmetry broken spontaneously by the background. In the 
Coulomb branch, we recover the leading contribution to the Dirac-Born-Infeld (DBI) action, exhibiting a
$S^5 \times AdS^5$ bulk geometry around a stack of branes. 
SUSY may be broken by compact extra dimensions $\cM^4 \times \cK$, 
leading to an induced  gravitational action on $\cM^4$ 
due to the trace--$U(1)$ sector. 
The one-loop effective action is UV finite on such backgrounds, and the UV/IR mixing is non-pathological.

\end{abstract}

\newpage
\tableofcontents

\section{Introduction}

One of the most fascinating ideas in recent years is the proposal that matrix models of Yang-Mills type, 
in particular certain supersymmetric models which have been put forward in string theory 
\cite{Ishibashi:1996xs,Banks:1996vh}, may 
provide a description for the quantum structure of space-time and geometry.
Even though these models are extremely simple, they contain all the required ingredients for 
a theory of fundamental interactions including gravity. In fact, they appear to reproduce 
at least the low-energy sector of string theory or supergravity in certain backgrounds. 
Here we  take these models as a starting point independent of string theory,
and focus on 4-dimensional brane solutions or backgrounds, considered as physical space(-time).
Due to  supersymmetry, these models should provide a well-defined quantum theory 
at least for 4-dimensional backgrounds. 
In fact, numerical evidence has been obtained recently  \cite{Kim:2011cr}
for the emergence of 3+1-dimensional space-time within the IKKT or IIB model. 
This provides renewed motivation to study the effective physics of  
4-dimensional backgrounds. 
The aim of this paper is to provide a better understanding of the 
quantum effective action on such backgrounds in the matrix model.

Solutions of the IIB model corresponding to flat branes are  well known, 
and are described by the Moyal-Weyl quantum plane $\R^{2n}_\theta$. 
Brane configurations with general geometry in the IIB model have also
been studied recently  \cite{Steinacker:2010rh}, clarifying 
the geometry and the low-energy physics of fluctuations around such backgrounds. 
At the semi-classical level, the geometry of these \nc $\,$ (NC) branes in the matrix model is clear: 
the fluctuations are governed by an effective metric which depends both on the 
Poisson structure and the embedding of the brane. 
However upon quantization, gravitational terms are induced on the brane. 
Thus quantum effects will play an important role for the 
dynamics of the brane geometry, i.e. for gravity. In fact, the relation with IIB supergravity in the bulk 
can only be seen at the quantum level, taking into account the one-loop effective action \cite{Ishibashi:1996xs}.
Similarly, the evidence for gravitons on branes obtained in 
\cite{Kitazawa:2006pj,Kitazawa:2005ih} relies on quantum effects.
 It is therefore very important to understand these quantum effects more explicitly. 

In principle, it is easy to write down the one-loop effective action in a formal, 
background-independent way \cite{Ishibashi:1996xs,Chepelev:1997av}.
This by itself is very interesting, and it was used to establish a 
relation with branes in IIB supergravity.
Furthermore, UV-finiteness on 4-dimensional backgrounds is obvious.
We start with a detailed derivation of this one-loop action, emphasizing the 
global $SO(9,1)$ resp. $SO(10)$ symmetry which is only broken spontaneously by the background.
The main subject of the present paper is then a detailed study of the 
1-loop action on more general backgrounds, both for non-Abelian gauge fields as well as for the 
trace-$U(1)$ sector governing the geometry.

Some steps in that program were taken in 
\cite{Blaschke:2010rr}, where the action induced by integrating out fermions in a matrix model was computed, 
and cast into a geometrical form. 
However, maximal supersymmetry is essential to make the matrix model finite in 
4 dimensions, without pathological UV/IR mixing and without any cutoff introduced by hand. 
On the other hand, SUSY must clearly be broken, 
otherwise no reasonably physics and no induced gravity will arise on the brane. 
We will show explicitly in \secref{sec:product-spaces} how SUSY can be broken dynamically within the IKKT model,
on suitable backgrounds with compactified extra dimensions $\cM^4 \times \cK$.
A gravitational action is then induced on the brane 
below the SUSY breaking scale, as suggested in \cite{Steinacker:2008ri}. 
This shows how realistic physics may emerge from the IKKT model, 
in the presence of refined compactification scenarios such as those 
discussed in \cite{Chatzistavrakidis:2011gs,Steinacker:2011wb}.
Without breaking $\cN=4$ SUSY, one only obtains a higher-derivative induced action for the trace-$U(1)$
sector, which we  study in \secref{sec:heatkernel-details}.

Besides the geometrical trace-$U(1)$ sector, the 1-loop contribution to the 
non-Abelian sector is also of great interest. 
Non-Abelian gauge fields arise on stacks of branes. For a single stack of $N$ coinciding branes, 
one obtains $SU(N)$ Super Yang-Mills (SYM) theory coupled to gravity.
In particular, one should expect the same quantum effects as in ordinary $\cN=4$ SYM theory.
This is studied in \secref{sec:nonabel} in two cases: first 
we compute the one-loop effective action for an unbroken massless gauge boson in the Coulomb phase. 
Following essentially standard steps adapted to the matrix model case, we obtain 
e.g. the well-known $\frac{F^4}{(\phi^2)^2}$ one-loop term. In fact we can largely maintain the
global $SO(10)$ symmetry while working on a generic curved brane. This allows
to obtain precisely the leading terms of the Dirac-Born-Infeld (DBI) action for a single 
brane at some distance of the stack of $N-1$ branes (corresponding to the Coulomb branch), 
and to identify the effective near-horizon $AdS^5 \times S^5$ 
geometry of such a background consistent with IIB supergravity. 
This illustrates how quantum effects may modify the
effective bulk geometry in the matrix model, and
provides additional evidence for the relation of the IIB matrix model with supergravity. 
In contrast to previous work on related issues such as 
\cite{Maldacena:1999mh,Alishahiha:1999ci,Berman:2000jw,Liu:2000ad}, 
our results are exclusively based on matrix model computations, without presupposing any relation with supergravity.
The mechanism is also reminiscent of holography
based on $\cN = 4$ SYM \cite{Maldacena:1997re}, but in fact non-commutativity allows a more direct 
understanding of the branes in the ``holographic'' bulk.

On the other hand, we emphasize that the effective gravity which is relevant for physics in a brane-world scenario
emerges on the brane, and it is not a simple reduction of the bulk gravity.
The effective metric on the brane involves the Poisson tensor in an essential way, which leads to 
novel mechanisms as shown in \cite{Rivelles:2002}. 
The present paper further clarifies the origin of the induced gravity terms on the brane, and 
takes some steps towards their explicit determination. Their precise form depends on the specific 
compactification, and more work is needed
to identify the appropriate compactifications and to understand the physics of the gravity sector.
This clarifies the relation of the emergent gravity viewpoint of the matrix model with string theory.

Finally, the 1-loop effective action in a background $\cM^4 \times \cK$ with fuzzy extra dimensions $\cK$
is studied in \secref{sec:product-spaces} both from the higher-dimensional geometry point of view, as well as from the 
4-dimensional point of view in terms of Kaluza-Klein modes originating from matrix-valued gauge fields. 
Special attention is paid to the 
origin of scales and symmetry breaking within the matrix model.

\section{The IKKT or IIB model}
\label{sec:IKKT-IIB}

We start with a general discussion of the IKKT (or type IIB) matrix model and its properties 
in the context of emergent gravity, and  explain how to compute the effective one-loop action. 
The IKKT or IIB model \cite{Ishibashi:1996xs} is defined by the following action
\begin{align}
S_{\rm IKKT}&=-(2\pi)^2\Tr\left(\co{X^a}{X^b}\co{X_a}{X_b}\,\, + \, 2\obar\Psi \gamma_a[X^a,\Psi] \right) 
\,,
\label{IKKT-MM}
\end{align}
where $X^a,\,\, a= 0,1,2,\ldots,9$ are Hermitian matrices,
$\Psi$ is a matrix-valued Majorana-Weyl spinor of $SO(9,1)$,
and the $\g_a$ form the corresponding Clifford algebra.
The model is obtained by dimensional reduction of the 10-dimensional $SU(N)$ Super-Yang-Mills theory
to a point, and taking the $N \to\infty$ limit. 
Indices are raised and lowered using the fixed background metric  $g_{ab} = \eta_{ab}$; later 
we will also discuss the Euclidean version where $g_{ab} = \delta_{ab}$.
This action is invariant under the following  symmetries:
\be
\begin{array}{lllll}
X^a \to U^{-1} X^a U\,,\quad  &\Psi \to U^{-1} \Psi U\,,\quad  &  U \in U(\cH)\,\quad  & \mbox{gauge invariance,}  \\
X^a \to \L(g)^a_b X^b\,,\quad  &\Psi_\a \to \tilde \pi(g)_\a^\b \Psi_\b\,,\quad  & g \in \widetilde{SO}(9,1)\,\; 
  & \mbox{Lorentz symmetry,}  \\
X^a \to X^a + c^a \one\,,\quad & \quad & c^a \in \R\,\quad  & \mbox{translational symmetry,}
\end{array}
\label{transl-inv}
\ee
where the tilde indicates the corresponding spin group, as well as $\cN=2$ matrix supersymmetry \cite{Ishibashi:1996xs}.
Here matrices are identified with operators on a separable Hilbert space $\cH$, and $U(\cH)$
denotes the group of unitary operators resp. matrices on $\cH$.

\subsection{The bosonic sector} 

We first focus on the bosonic part of the IKKT model, given by
\begin{align}
S_{\rm YM} &=-(2\pi)^2\Tr\left(\co{X^a}{X^b}\co{X_a}{X_b}\right)
\,.
\label{bosonic-MM}
\end{align}
We want to compute the effective action, and for this purpose we will employ the background field method 
(cf. e.g.~\cite{Peskin:1995,Weinberg:1980wa,Khoze:2000sy,Chepelev:1997av}). 
Hence we split the matrices into background $X^a$ and a fluctuating part $Y^a$, 
\be
X^a\to X^a+Y^a \,. 
\ee
For a given background $X^a$, there is a gauge symmetry
\begin{align}
 Y^a \to  
 Y^a + U [X^a + Y^a, U^{-1}]
\,, \label{eq:Y-gaugesymm}
\end{align}
which we fix using the gauge-fixing function $G[Y] = i[X^a,Y_a]$.
We also need to add ghosts and antighosts $c$ resp. $\bc$, and together with the gauge fixing the additional terms read
\begin{align}
S_{\rm g.f.} + S_{\rm ghost} &=4(2\pi)^2\Tr\,s\left(\bc\, G[Y]-\frac{\a}{2}\bc b\right) \nn\\
&=4(2\pi)^2\Tr\left(b\,i\co{X^a}{Y_a}-\frac{\a}{2}b^2-\bc\,i\co{X^a}{sY_a}\right)\,,
\end{align}
where $b$ is a multiplier field fixing the gauge, and the BRST transformations associated to the gauge symmetry \eqref{eq:Y-gaugesymm} are given by 
\begin{align}
sY_a&=i\co{X_a+Y_a}{c}\,, && s\bc=b\,, && sb=0 
\,, \nn\\
sc&=ic^2\,, && s^2\varphi=0 &\forall\varphi\in&\{Y_a,c,\bc,b\} 
\,. 
\end{align}
Choosing $\a=1$ and eliminating $b$ using its e.o.m. 
\begin{align}
\var{S}{b}=4(2\pi)^2\left(i\co{X^a}{Y_a}-b\right)=0
\,, 
\end{align}
we arrive at
\begin{align}
S_{\rm g.f.} + S_{\rm ghost} &= - 2(2\pi)^2\Tr\left(\co{X^a}{Y_a}\co{X^b}{Y_b}-2\bc\co{X^a}{\co{X_a+Y_a}{c}}\right)
\,. \label{eq:action_gf-part}
\end{align}
In a one-loop computation of the effective action in $X^a$, one keeps only the quadratic terms in the 
fluctuations $Y$ and discards linear and higher order terms \cite{Peskin:1995}. 
For the gauge invariant part of the action \eqref{bosonic-MM} this yields
\begin{align}
S_Y&=2(2\pi)^2\Tr\left(Y^a \co{X^b}{\co{X_b}{Y_a}} - Y^a \co{X^b}{\co{X_a}{Y_b}}
 + Y^a[[X_a,X^b],Y_b]\right) 
. \label{eq:action_inv-part}
\end{align}
Combining \eqref{eq:action_inv-part} with \eqref{eq:action_gf-part}, one arrives at the following quadratic action
\begin{align}
S_{\rm quad} 
&= 2(2\pi)^2\Tr\left(Y^a \co{X^b}{\co{X_b}{Y_a}}+2iY_a \co{\Th^{ab}}{Y_b}+2\bc\co{X^a}{\co{X_a}{c}}\right)  \nn\\
&= 2(2\pi)^2\Tr\left(Y^a (\Box \d^{ab} + 2i[\Th^{ab},.]) Y_b +2 \bc\Box c \right)  \nn\\
&= 2(2\pi)^2\Tr\left(Y^a (\Box  + \Sigma^{(Y)}_{rs}[\Th^{rs},.]))^a_b Y^b +2 \bc\Box c \right)  
\,, \label{quad-action}
\end{align}
where we define
\begin{align}
\Th^{ab} &=-i\co{X^a}{X^b} \,, &
(\Sigma_{ab}^{(Y)})^c_d &= i(\d^c_a g_{bd} - \d^c_b g_{ad})\, , \nn\\
\Box \phi &= [X^a,[X_a,\phi]]\, .
\end{align}
Note that
when the background is not fixed, the combined action is invariant under the background gauge transformations
\be
 X^a \to  U X^a U^{-1}\,, \qquad Y^a \to U Y^a U^{-1}\,,  \qquad c \to U c U^{-1}
\,. 
\ee
It follows that the one-loop effective action induced by the bosonic matrices\footnote{Note that 
from the NC gauge theory point of view, they comprises 
both gauge fields and scalar fields.} $\Gamma_{{\rm 1-loop}}[X]$  given by
\bea
\Gamma_{{\rm 1-loop}}[X] = -\inv2 \Tr \log(\Box  + \Sigma^{(Y)}_{rs}[\Th^{rs},.]) + \Tr \log(\Box)  
\eea
also enjoys this fundamental gauge invariance, 
provided the integrals over the fluctuations $Y$ and $c$ are regularized such that this symmetry is preserved. 
In analogy to the fermionic integral  computed in~\cite{Blaschke:2010rr}, this can be achieved 
using a Schwinger parametrization as follows:
\be
\Gamma_{{\rm 1-loop,L}}[X] := \Gamma_{{\rm 1-loop,L}}^Y[X] +  \Gamma_{{\rm 1-loop,L}}^c[X]
\,, 
\ee
where
\begin{align}
\Gamma_{{\rm 1-loop,L}}^Y[X] &:= \frac 12 \Tr \int_0^\infty \frac {d\a}{\a} 
    e^{-\a(\Box + \Sigma^{(Y)}_{ab}[\Th^{ab},.])}e^{-\frac 1{\a L^2}} \,, \nn\\
\Gamma_{{\rm 1-loop,L}}^c[X] &:= -\Tr \int_0^\infty \frac {d\a}{\a} e^{-\a\Box } e^{-\frac 1{\a L^2}}  
\,. \label{Gamma-schwinger}
\end{align}
Here $L$ is a cutoff of dimension ``length'', which 
essentially sets a lower limit $\a >\frac 1L$ for the $\a$ integral.
This amounts to a UV cutoff
\be
\L := \LNC^2 L 
\label{eq:def-cutoff}
\ee
of dimension (length)$^{-1}$ in a background with NC scale $\LNC$, where $X^a$ is considered to have dimension length.
Note that the two traces in \eq{Gamma-schwinger} are different. By construction, both contributions are gauge invariant
and satisfy a scaling relation
\begin{align}
\Gamma_{{\rm 1-loop,L}}[X] &= \Gamma_{{\rm 1-loop,L}}[U X U^{-1}]\, , \nn\\
\Gamma_{{\rm 1-loop,c L}}[cX] &= \Gamma_{{\rm 1-loop,L}}[X]
\,. 
\end{align}
We also expect invariance under $SO(D)$ as in \cite{Blaschke:2010rr}.
However, we will see that no such regularization is needed in the maximally supersymmetric
IKKT model.

Finally, it is straightforward to include also e.g. a mass term $\Tr\, m^2 X_a X^a$ or a 
cubic term $\Tr\, i C_{abc} X^a [X^b, X^c]$  into the formalism. 
Then the quadratic action \eq{quad-action} becomes
\begin{align}
S_{\rm quad} &= 2(2\pi)^2\Tr\left(Y^a 
 \big(\Box +m^2 + \Sigma^{(Y)}_{rs}[\Th^{rs},.]\big)_{ab} Y^b  + Y^a i C_{abc} [X^c,Y^b] +2 \bc\Box c \right)  
\label{quad-action-m}
\end{align}
for totally antisymmetric $C_{abc}$.

\subsection{The fermionic sector} 

\paragraph{Minkowski case.} 
For Dirac fermions, the functional determinant can be simply evaluated as
\begin{align}
\int d\obar\Psi d\Psi e^{i\Tr \obar\Psi \gamma_a[X^a,\Psi]} &= \det(i\Gamma_0\slashed{D}) 
= \sqrt{\det \slashed{D}^2}
\,. 
\end{align}
However, the fermions in the IKKT model formulated in $9+1$ dimensional Minkowski space-time are (matrix-valued) Majorana-Weyl (MW) spinors $\Psi$ of $SO(1,9)$, 
in particular
\be
\Psi_C = \cC \obar\Psi^T = \cC \gamma_0^T\Psi^\star = \Psi
\,, 
\ee
where $\gamma_a^T = \cC \gamma_a \cC^{-1}$.
This means that in a MW basis, the spinor entries are Hermitian (rather than 
general complex) matrices (cf.~\cite{Nishimura:2001sx,vanNieuwenhuizen:1996tv,Wetterich:2010ni}). 
In 9+1 dimensions, 
$\cC = \cC^{-1T}$ anticommutes with the chirality operator $\gamma = i\gamma_0 \ldots \gamma_9$, 
so that the symmetric matrix
\be
 \tilde \gamma_a = \cC \gamma_a
\label{gamma-AS}
\ee
 is well-defined for Weyl spinors. 
Using $\obar\Psi = \Psi^T \cC^{-1 T} $, the fermionic action can be written as
\begin{align}
 \Tr \obar\Psi \gamma_a[X^a,\Psi] &= \Tr \Psi^T \cC^{-1T} \gamma_a[X^a,\Psi] 
= \Tr \Psi_\a \tilde\gamma_a^{\a\b} [X^a,\Psi_\b] 
\,, 
\end{align}
where we assume that $\Psi$ are 16-dimensional Weyl spinors,
or equivalently 
\begin{align}
&\Tr \obar\Psi \gamma_a[X^a,\Psi] = \Tr \Psi^T \cC \slashed{D}P_+\Psi\,, && P_\pm = \frac 12(1\pm\gamma) 
\,, 
\end{align}
for 32-component Dirac spinors.
Then the Grassmann integral over the MW spinors yields 
\begin{align}
e^{i\Gamma^\psi[X]} &:=\int d\Psi e^{i\Tr \obar\Psi \gamma_a[X^a,\Psi]} 
  = {\rm Pfaff}(\tilde\gamma_a^{\a\b} [X^a,.]) \nn\\
&= \pm \sqrt{\det \tilde\gamma_a^{\a\b} [X^a,.]} = \pm\sqrt{\det(\cC\slashed{D}_+)}
\,, \label{eq:Grassmann-integral-MW-spinors}
\end{align}
where\footnote{The transpose in $\Psi^T$ refers only to the 10 spinor indices.} 
$\cC\slashed{D}_+$ denotes $\cC\slashed{D}$ acting on the positive chirality spinors,
to be specific.

\paragraph{Euclidean case.} 

Since the Pfaffian makes sense for any (complex) anti-symmetric matrix, one can use 
 \eqnref{eq:Grassmann-integral-MW-spinors} to define the Wick-rotated fermionic induced action 
$\Gamma_E^\psi[X]$ also in the Euclidean 
case (at least for finite-dimensional matrices), by replacing $\gamma_0 \to i \gamma_{10}$ 
(see Ref.~\cite{vanNieuwenhuizen:1996tv,Wetterich:2010ni} for further details on Wick rotations in the context of spinors). 
However, then $\tilde\gamma_a^{\a\b} [X^a,]$ is in general not a Hermitian operator for chiral 
$SO(10)$ spinors, and
the effective action has both real and imaginary contributions. 
The real part of the action can be extracted from
\begin{align}
\det \((\cC \slashed{D})^\dagger \cC \slashed{D}_+\) = \det(\slashed{D}^2_+)
 = e^{-2\Re(\Gamma_E^\psi[X]) } \,,
\end{align}
which is real because $\slashed{D}^2$ is a Hermitian operator. 
Therefore we can write
\begin{align}
 \Gamma_E^\psi[X] \, &= \Gamma_E^{\psi,\rm real}[X] \, + i \Gamma_E^{\psi,\rm imag}[X] \, \nn\\
&= \,\,- \frac 14\Tr \log (\slashed{D}^2_+)\,\, + i \Gamma_{WZ}
\,. 
\end{align}
The imaginary part is the Wess-Zumino contribution.
It is invariant under $SO(10)$ (but not ``locally''), and 
incorporates the anomaly contribution due to the  integrated out fermions.
For a more detailed discussion we refer to 
\cite{D'Hoker:1984ph,Ishibashi:1996xs,Tseytlin:1999tp,Belyaev:2011dg}. 

For the real part of the effective action,
it is useful to introduce a representation using Schwinger parameters, based on the  
identity
\be
\int_0^\infty\frac{d\a}{\a}(e^{-a X} - e^{-a Y}) = \log\frac{Y}{X}
\,. 
\ee
Noting that $\slashed{D}^2 = \Box + \Sigma^{(\psi)}_{ab}[\Th^{ab},.]$, we obtain
\begin{align}
\Gamma_E^{\psi,\rm real}[X] \, = \,\,\inv{2}\Tr\int\limits_0^\infty\frac{d\a}{\a}
\,e^{-\a(\Box +\Sigma_{ab}^{(\psi)}[\Theta^{ab},.])} ,  
 \label{TrLog-id-fermions}
\end{align} 
where $\Sigma^{(\psi)}_{ab}$ denotes the chiral representation on Weyl spinors as given below.
The fermionic induced action can also be regularized by introducing
an UV cutoff term $e^{-\inv{\a L^2}}$ as in \eq{Gamma-schwinger}.
However this will not be necessary 
in the maximally supersymmetric case under consideration here.
In the following we will simply replace 
$\Gamma_E^{\psi}$ by $\Gamma_E^{\psi,\rm real}$ dropping $\Gamma_{WZ}$, which would not 
modify our explicit computations below. The  Wess-Zumino contribution is expected to be finite
and $SO(10)$ invariant in a ``non-local'' manner, 
and would deserve a detailed investigation extending \cite{Tseytlin:1999tp}.

\subsection{1-loop  effective action for the IKKT model}

Now consider the complete model including bosons and fermions. 
First, we introduce a common notation which works for all fields.
Let 
\begin{align}
\begin{array}{rlcl}(\Sigma_{ab}^{(\psi)})^\a_\b &= \frac i4 [\gamma_a,\gamma_b]^\a_\b \,, && \mbox{fermions,}  \\
                      (\Sigma_{ab}^{(Y)})^c_d &= i(\d^c_a g_{bd} - \d^c_b g_{ad}) \,, && \mbox{bosonic matrices,} \\
                 \Sigma_{ab}^{(c)} &= 0 \,, && \mbox{ghosts,}
                    \end{array}  
\end{align}
be the generators of $SO(D)$ on the spinor and vector irreducible representations 
($[\gamma_a,\gamma_b]$ is understood to be the irreducible chiral representation).  
To evaluate the one-loop effective action explicitly, we will need the traces $\tr \Sigma_A \ldots \Sigma_C $
where $\Sigma_A \equiv \Sigma_{ab}$ denotes the generators of $SO(D)$ in the vector or spinor representation.
These are computed in \appref{app:invariants}, with the results 
\begin{align}
\tr \Sigma_A  &= 0\,,  &
\tr \Sigma_{A} \Sigma_B &=   \frac{2\tr \one}{D(D-1)}  C^{(2)}\, g_{AB}\,,  \nn\\
\tr \Sigma_{A} \Sigma_B \Sigma_C &=   i \frac{\tr \one}{D(D-1)} C^{(2)}\, f_{ABC} 
\,. 
\end{align}
Taking into account the results of the previous section, we will now switch to the Euclidean case, 
and drop the (imaginary) Wess-Zumino term in the effective action. 
The real part of the 1-loop contribution for the  IKKT model can therefore be written as
\begin{align}
\vG:=\Gamma^{\rm real}_{{\rm 1-loop}}[X] &= - \frac 12 \Tr \int\limits_0^\infty \frac {d\a}{\a} 
   \Big( e^{-\a(\Box  + \Sigma^{(Y)}_{ab}[\Th^{ab},.])}
  - \frac 12 e^{-\a(\Box  + \Sigma^{(\psi)}_{ab}[\Th^{ab},.]) }
 - 2 e^{-\a\Box}  \Big) 
\,, \label{Gamma-schwinger-susy}
\end{align}
which is manifestly $SO(10)$ invariant\footnote{The $SO(3,1) \times SO(D-4)$ 
structure was pointed out also in \cite{Khoze:2000sy}, 
but the $SO(10)$ covariant form apparently has not been used up to now.}. 
As pointed out before, there may be an additional imaginary contribution to the action
for some backgrounds, which can be 
interpreted as Wess-Zumino term related to a global anomaly 
of the chiral $SU(4)$ R-symmetry. An analogous formula applies for the $D$-dimensional reduced model.

As a first check, we note that $\vG[X]$ vanishes 
in the free case (i.e. for an unperturbed $\R^4_\theta$ background with $\Theta^{ab} \sim \one_\cH$), 
provided the  $D-2$ effective degrees of freedom from bosons - ghosts
cancel with the fermions. This holds in the $D=10$ IKKT model for Majorana-Weyl fermions, and also in the 
$D=6$ model for Weyl fermions in the $(4)$ of $SO(6)$. The latter corresponds to $\cN=2$ SUSY Yang-Mills
from a 4-dimensional point of view. 
These cancellations hold for any background where $[\Theta^{ab},.] = 0$. 
Adding perturbations around $\R^4_\theta$, the $\a$ integral becomes divergent and must be regularized as in 
\eqnref{TrLog-id-fermions}  in the
$D\neq10$ case, \emph{except} for the 
maximally supersymmetric $D=10$ case which is (believed to be) 
perturbatively finite\footnote{This is a strong hint that 4-dimensional 
backgrounds in the IKKT model are stable and preferred.} to all loops.
One-loop finiteness will become obvious in the following.

The above formula \eq{Gamma-schwinger-susy} is valid for arbitrary backgrounds.
It suggests the following expansion:
\begin{align}
\vG[X]\! &:= - \frac 12 \Tr \int_0^\infty \frac {d\a}{\a} 
   \Big( e^{-\a(\Box  + \Sigma^{(Y)}_{ab}[\Th^{ab},.])}
  - \frac 12 e^{-\a(\Box  + \Sigma^{(\psi)}_{ab}[\Th^{ab},.]) }
 - 2 e^{-\a\Box}  \Big) \nn\\
&= \frac 12 \Tr \(\log(\Box  + \Sigma^{(Y)}_{ab}[\Th^{ab},.])
-\frac 12 \log(\Box  + \Sigma^{(\psi)}_{ab}[\Th^{ab},.])
- 2 \log \Box\)   \nn\\
&= \frac 12 \Tr \(\log(\one  + \Sigma^{(Y)}_{ab}\Box^{-1}[\Th^{ab},.])
-\frac 12 \log(\one  + \Sigma^{(\psi)}_{ab}\Box^{-1}[\Th^{ab},.])\) \nn\\
 &= \frac 12 \Tr \Bigg(\sum_{n>0} \frac{(-1)^{n+1}}n \Big((\Sigma^{(Y)}_{ab}\Box^{-1}[\Th^{ab},.])^n 
  \, -\frac 12 (\Sigma^{(\psi)}_{ab}\Box^{-1}[\Th^{ab},.])^n \Big)  \Bigg) \nn\\
 &= \frac 12 \Tr \Bigg(\!\! -\frac 14 (\Sigma^{(Y)}_{ab} \Box^{-1}[\Th^{ab},.] )^4 
 +\frac 18 (\Sigma^{(\psi)}_{ab} \Box^{-1}[\Th^{ab},.])^4 \,\, +  \cO(\Box^{-1}[\Th^{ab},.])^5 \! \Bigg) .
\label{Gamma-IKKT}
\end{align}
Note that the first terms with $\Sigma, \, \Sigma^2\,$ and $\Sigma^3$ cancel,
as was observed in \cite{Ishibashi:1996xs}. 
Moreover, the $SO(10)$ resp. $SO(9,1)$ symmetry is still manifest. 
These statements are independent of any specific background, and reflect the maximal SUSY of the model.
Therefore the traces  behave
as $\int\! d^{2n} p\, \frac 1{p^8}$ in the UV on $2n$-dimensional backgrounds, which is convergent
for $2n < 8$.
In particular, the model is one-loop finite\footnote{One-loop finiteness
holds even on 6D backgrounds, but not for higher loops. This is consistent with well-known
results \cite{Green:1982sw,Fradkin:1983jc,Howe:1983jm} for SYM theory.}  on 4-dimensional backgrounds.
There will be oscillating contributions from phase factors due to the non-commutativity,
however they are harmless here because the traces are absolutely convergent.
Hence the model is free of pathological UV/IR mixing, at least up to one loop
(however there might be standard IR issues, as discussed below).
This expression is  background independent, and applies both to the Abelian and  to the non-Abelian case.

Consider the first non-vanishing term 
\begin{align}
&\vG[X] = \frac 18 \Tr_\algA \bigg(\! \tr\Big(\!-\Sigma^{(Y)}_{A} \ldots \Sigma^{(Y)}_{D} + \tinv{2}\Sigma^{(\psi)}_{A} \ldots \Sigma^{(\psi)}_{D} \Big)
  \Box^{-1}[\Th^{A},\Box^{-1}[\Th^{B},\Box^{-1}[\Th^{C},\Box^{-1}[\Th^{D},.]]]]  \nn\\
&\qquad\hspace*{1.7cm}  + \ldots\!\! \bigg)  
\,. \label{Gamma-IKKT-4}
\end{align}
Here a sum over $\mso(10)$  indices $A \equiv (a,b);\,\,a<b$ is understood, and $\Tr_\algA$ denotes the 
trace over the space of (possibly $\msu(N)$ valued) 
functions on the background brane without spinor indices.  
It is interesting to note that in contrast to the non-supersymmetric case  \cite{Blaschke:2010rr},
this term is proportional to $(\Theta^{ab})^4$, and the commutator structure will lead to 
a complicated momentum dependence for the $U(1)$ sector. 
The results of the non-SUSY case 
will re-surface in the case of supersymmetry breaking in \secref{sec:product-spaces}.
To proceed, we need $\tr (\Sigma_{A}\ldots\Sigma_{D})$ 
for both vector and spinor representations, which is given in \eq{four-sigma-trace}.

\subsection{Non-commutative branes and effective geometry}

One of the particularly interesting features of the IKKT model \eqref{IKKT-MM}
 is that it incorporates a quantum theory of gravity in an emergent way --- for a detailed review
see~\cite{Steinacker:2010rh,Blaschke:2010rg,Blaschke:2010qj} and references therein.
The key is to consider the matrices $X^a$ as ``quantized  embedding functions''
$X^a\sim x^a:\,\,\cM \hookrightarrow \R^{10}$ in the semiclassical limit, where $x^a$ are the 
Cartesian coordinate functions of $\R^{10}$. 
In particular, $U(1)$--valued deformations  of the Moyal-Weyl quantum plane
as in \eq{eq:general-X} can be interpreted geometrically 
in terms of deformed embedded branes \cite{Steinacker:2008ri}
\be
X^a = \begin{pmatrix}
       X^\mu \\ \phi^i(X^\mu)
      \end{pmatrix} : \quad \cM^4 \hookrightarrow \R^{10} 
\,. \label{eq:matrix-splitting}
\ee
Here $X^\m\, , \m = 1,\ldots,4$ are considered as independent quantized coordinate functions
satisfying generic commutation relations $[X^\mu,X^\nu] = i \theta^{\mu\nu}(X)$ 
(which can be interpreted as a quantized Poisson structure on $\cM^4$), 
while the $\phi^i(X^\mu)$ are  functions of these coordinates.
Then the matrix model action \eq{IKKT-MM} can be viewed as describing
scalars, gauge fields and fermions propagating on $\cM^4\subset \R^{10}$ with effective metric~\cite{Steinacker:2007dq}
\begin{align}
G^{\mu\nu} &= e^{-\s}\theta^{\mu\mu'}\theta^{\nu\nu'} g_{\mu'\nu'} \,,  &
g_{\mu\nu} &= \del_\mu x^a \del_\nu x^b g_{ab} \,, &
e^{-\s}&\equiv \frac{\sqrt{\det\th^{-1}_{\m\n}}}{\sqrt{\det G_{\r\s}}}
\,, \label{eq:def-Gges}
\end{align}
in the semi-classical limit where $X^a\sim x^a$ and $\theta^{\mu\nu}(X) \sim \theta^{\mu\nu}(x)$. 
To introduce dimensions, we 
can use the fixed background metric $g_{ab} = \delta_{ab}$ resp. $g_{ab} = \eta_{ab}$ of the matrix model
to define an (unphysical) scale $\L_0$,  such that $\L_0=1$ for $|g_{ab}| = 1$. 
In that sense, $X^a$ has dimension length, 
and $\theta^{\mu\nu}$ encodes the (physical) non-commutativity scale
\be
\bLNC^4 = \det{\th^{-1}_{\m\n}} \quad =: \L_0^4\, e^{-\sigma} \,,
\ee
which may depend on $x$. 

The effective geometry becomes clear in the following example: 
The gauge invariant kinetic term of a test particle modeled by a scalar field $\phi$ has the form
\begin{align}
S[\phi]&=-\Tr\co{X^a}{\Phi}\co{X^b}{\Phi}\eta_{ab} \,
\sim\intx\sqrt{\det\th^{-1}}\th^{\m\n}\pa_\m x^a\pa_\n\phi\th^{\r\s}\pa_\r x^b\pa_\s\phi\eta_{ab} \nn\\
&=\intx\sqrt{\det G}G^{\n\s}\pa_\n\phi\pa_\s\phi
\,.
\end{align}
Hence, the action in the semiclassical limit is that of a scalar field propagating in curved space-time with metric $G^{\m\n}$. 

This geometrical view of the matrix model is much more transparent 
than the interpretation in terms of NC $U(1)$ gauge theory, while the latter is 
more suitable for perturbative computations.

\subsection{Background expansion}

To obtain explicit expressions for the one-loop effective action,
we will consider a background of slowly varying gauge and scalar fields around $\R^4_\theta$,
\be
X^a = \begin{pmatrix}
 \bar X^\mu \\ 0
\end{pmatrix} \, + \, 
\begin{pmatrix}
 \cA^\mu  \\   \phi^i
\end{pmatrix} \,,
\label{eq:general-X}
\ee
and treat $\cA^\mu = -\bar\theta^{\mu\nu}A_\nu(\bar X)$ and $\phi^i = \phi^i(\bar X)$ as gauge fields resp. scalar
fields on $\R^4_\th$.
Then
\begin{align}
[X^a,[X_a,\Psi]] 
&= \bar\Box \Psi + \d_{\mu\nu}([\bar X^\mu,[\cA^\nu,\Psi]] +[\cA^\mu,[\bar X^\nu,\Psi]] + [\cA^\mu,[\cA^\nu,\Psi]]) 
 + \d_{ij}[\phi^i,[\phi^j,\Psi]] \nn\\ 
&= \bar\Box \Psi - \frac{i\bG^{\m\n}}{\bLNC^4}
\left(2[A_{\mu},\del_{\nu}\Psi] + [\del_{\mu}A_{\nu},\Psi] + i[A_{\mu},[A_{\nu},\Psi]]\right)
+   [\phi^i,[\phi_i,\Psi]]  
\,, \label{eq:def-V}
\end{align}
where 
\begin{align}
\bar\Box\Psi=\co{X^\m}{\co{X_\m}{\Psi}}=-\bLNC^{-4}\bG^{\m\n}\pa_\m\pa_\n\Psi
\,
\end{align}
is the free Laplace operator on $\R^4_\th$. 
The one-loop effective action \eq{Gamma-IKKT} involves commutators with the generalized field strength defined by 
\begin{align}
\Theta^{ab} &= \bar\Theta^{ab} + \cF^{ab} 
\,, 
\end{align}
which satisfies $[\Theta,.] = [\cF,.]$. 
Note that the $10 \times 10$ matrix
\begin{align}
\Theta^{ab} &= \begin{pmatrix}
                             -\theta^{\mu\mu'}\theta^{\nu\nu'} (\theta^{-1} + F)_{\mu\nu}  & \theta^{\mu\mu'} D_{\mu'}\phi^i \\
                 - \theta^{\nu\nu'} D_{\nu'}\phi^j &  [\phi^i,\phi^j]    \end{pmatrix} 
\label{Theta-explicit-4D}
\end{align}
forms an irreducible representation of $SO(9,1)$ resp. $SO(10)$. Its non-trivial part
$\cF^{ab}$ continues to make sense in the commutative case 
(upon absorbing  $\theta^{\mu\nu}$ and $\bLNC^2$ in $\Sigma^{ab}$ resp. $\phi^i$), where it
decomposes into  field strength and covariant derivatives transforming under
$SO(3,1) \times SO(6)$. Hence gauge fields and scalar fields are related by
$SO(10)$  in the {\nc} case, which is very remarkable from the field-theoretical point of view.
The effective action is expected to respect this $SO(10)$ invariance,
which is broken spontaneously by the background and hence non-linearly realized on the fields; 
for the fermionic induced action this was verified in \cite{Blaschke:2010rr}.

These formulas apply both in the Abelian and the non-Abelian case, however the
physical content is very different.
In the Abelian case, the commutators in \eqnref{Gamma-IKKT-4} result from Poisson brackets
\be
[\cF^{ab}(k),\phi(p)] \sim i \{\cF(k),\phi(p)\} =   \cO(kp) 
\,. 
\ee
Hence non-commutativity is essential here, but does not lead to 
pathological  UV/IR mixing due to maximal SUSY.
The leading term in a low-energy (momentum) expansion will arise from
$[\cF^{ab},.]^4 = \cO(k^4 (F^4,(\del\phi)^4))$, corresponding to
local as well as non-local terms at $\cO(k^4)$. 
In particular, no potential term is induced.

In the Non-Abelian case,
\be
[\cF^{ab}(k),\phi(p)] =  \cO(1) 
\ee
for the $\msu(N)$-valued fields,
leading to a non-vanishing potential at $k\to 0$. On a flat $\R^4_\th$ background, the $SU(N)$
sector essentially reduces to $\cN=4$ SYM gauge theory at low energies. 
Indeed, we have
\be
\Box^{-1} \,\, \stackrel{k=0}{\to} \,\,  \bLNC^{4}\frac 1{\dott{p}{p} +  [A^\mu,[A_\mu,.]] + [\phi^i,[\phi^i,.]] } 
\,, 
\ee
where
\be
\dott{p}{p} = \bar G^{\mu\nu} p_\mu  p_\nu .
\ee
Thus the 1-loop low energy effective action for the $SU(N)$ sector
has precisely the same form as for the $\cN=4$ SYM theory in the commutative case,
consisting of terms of the structure
\begin{align}
\vG[A,\phi] &= \int\frac{d^4 p}{(2\pi\bLNC)^2} 
\Tr\sum \Big(\bLNC^{4}\inv{\dott{p}{p} +  [A^\mu,[A_\mu,.]] + [\phi^i,[\phi^i,.]]} \Sigma_{ab}[\Theta^{ab},.]]\Big)^k .
\end{align}
Turning on also non-trivial trace-$U(1)$ components, this $SU(N)$ action is  coupled 
to non-trivial background geometries, 
as expected from the emergent gravity point of view \cite{Steinacker:2010rh}.
This will be discussed in \secref{sec:nonabel}.

\section{Evaluation of the one-loop effective action: Abelian case}
\label{sec:heatkernel-details}

Consider first the Abelian case. 
Assuming a background given by a small perturbation of $\R^4_\th$ as above, we can expand
\be
\Box^{-1} = \bar\Box^{-1}\(1-\bar\Box^{-1}V+ (\bar\Box^{-1}V)^2 - \ldots\)
\label{prop-expand}
\ee
where $\Box = \bar\Box + V$. 
In contrast to the non-supersymmetric case \cite{Blaschke:2010rr}, the leading non-trivial contributions 
are obtained here by replacing $\Box \to \bar\Box$, so that we can neglect the 
sub-leading terms in \eq{prop-expand} here. 
We  introduce Schwinger parameters for the propagators
\begin{align}
\Box^{-1}&\approx \bar\Box^{-1} 
= -\int_0^\infty d\a\, e^{-\a\bar\Box} 
= -\bLNC^4\int_0^\infty d\bar\a\, e^{-\bar\a\bLNC^4\bar\Box}
\,
\end{align}
where $\bar\a = \bLNC^{-4} \a$, and use a basis of plane waves such that 
commutators can be expressed in terms of sines, cf. \cite{Blaschke:2010rr}. 
Then the matrix element for the operator in \eqnref{Gamma-IKKT-4} can be written as follows:
\begin{align}
\big\langle q|\widetilde{V}|p \big\rangle &\equiv \big\langle q| \bar\Box^{-1}[\Th^{A},\bar\Box^{-1}[\Th^{B},\bar\Box^{-1}[\Th^{C},\bar\Box^{-1}[\Th^{D},.]]]] |p \big\rangle \nn\\
&=\int\!\!\frac{d^4 q}{(2\pi\bLNC^2)^2}\Psi(q)\int\!\!\frac{d^4pd^4k_1d^4k_2d^4k_3}{2^6\pi^{10}}\int_{0}^{\infty}\!d\bar\a_{1-4}\Th^A(q-p-k_1-k_2-k_3)\Th^B(k_3) \nn\\
&\quad\times \Th^C(k_2)\Th^D(k_1)\exp\left[-\bar\a_1\dott{(k_1+p)}{(k_1+p)}-\bar\a_2\dott{(k_1+k_2+p)}{(k_1+k_2+p)}\right] \nn\\
&\quad\times \exp\left[-\bar\a_3\dott{(k_1+k_2+k_3+p)}{(k_1+k_2+k_3+p)}-\bar\a_4(\dott{q}{q})\right]\Sin{k_1\th p} \nn\\
&\quad\times \Sin{k_2\th(k_1+p)}\Sin{k_3\th(k_1+k_2+p)}\Sin{q\th(k_1+k_2+k_3+p)}\Psi(p)
\,, \label{eq:V-matrix-element}
\end{align}
and hence
\begin{align}
\Tr\widetilde{V}&=\int\!\!\frac{d^4pd^4k_1d^4k_2d^4k_3}{2^6\pi^{10}}\int_{0}^{\infty}\!d\bar\a_{1-4}\Th^A(-k_1-k_2-k_3)\Th^B(k_3)\Th^C(k_2)\Th^D(k_1) \nn\\
&\quad\times \exp\left[-\bar\a_1\dott{(k_1+p)}{(k_1+p)}-\bar\a_2\dott{(k_1+k_2+p)}{(k_1+k_2+p)}\right] \nn\\
&\quad\times \exp\left[-\bar\a_3\dott{(k_1+k_2+k_3+p)}{(k_1+k_2+k_3+p)}-\bar\a_4(\dott{p}{p})\right]\Sin{k_1\th p} \nn\\
&\quad\times \Sin{k_2\th(k_1+p)}\Sin{k_3\th(k_1+k_2+p)}\Sin{p\th(k_1+k_2+k_3)}
\,. \label{O4-trace-1}
\end{align}

\paragraph{Note on the IR problem.}

In the context of massless gauge fields one typically encounters an infrared (IR) problem. 
More specifically, if the external momenta $k_i$ are set to zero, then the integral over the internal loop 
momentum diverges at $p = 0$, or equivalently the integral over the Schwinger parameter $\a$ diverges at
$\a \to \infty$. In the present case, this does not seem to happen. The reason is that the 
coupling is proportional to the $\sin(p \theta k)$ which vanishes at $k=0$. The 
price to pay is a complicated non-local form of the effective action.

To analyze the IR behaviour in more detail, note that the trigonometric functions in \eq{O4-trace-1}  
contribute a factor of order $\cO(k,p)^8$.
Upon carrying out the Gaussian integration over $p$, 
every sine-factor (which arises from a commutator in the vertex \eqref{eq:V-matrix-element}) 
leads to an $\cO(k^2)$ term. These then take care of the $\frac 1{k^2}$ terms from the propagators
(which arise from the integration over the Schwinger parameters here). 
Therefore there will be no IR divergences. In fact,
the above argument implies that the leading term in the effective action is $\cO(k^4)$, 
multiplying four $\Theta^{ab}$ factors. Since the $\Theta^{ab}$ can be interpreted 
either in terms of $(F,\del\phi^i)$ or in terms of 
a generalized vielbein \cite{Steinacker:2008ri},
this corresponds to a higher-derivative geometrical term whose physical significance is unclear.

We conclude that the 1-loop integral is both UV and IR finite in the Abelian 
case or in the trace--$U(1)$ sector, with complicated non-local low-energy behaviour starting
at $\cO(k^4 (\Theta^{ab})^4)$. 
More specifically, we will obtain terms which are homogeneous functions of order 4 in the $k_i$, 
as well as ``quasi-homogeneous'' terms 
with logarithmic behaviour.

\paragraph{Gaussian integration.}

To proceed with the explicit loop integral, we first need to integrate over $p$.
Since none of the $\Th^A$ depend on $p$,
the Gaussian integral over $p$ can be carried out trivially using
\begin{align}
\int\! d^4 p \exp(-\l \dott{p}{p} -2 \l \dott{q}{p}) e^{i k \th p} 
&= \frac{\pi^2}{\l^2} \exp( \l \dott{q}{q}- \frac{\dott{\tilde k}{\tilde k}}{4\l}) e^{-i k\th q} , \nn\\
\int\! d^4 p \exp(-\l \dott{p}{p} -2 \l \dott{q}{p}) (\sum_a g_a (k_i)  e^{i k_a \th p})
&= \frac{\pi^2}{\l^2}  \exp(\l \dott{q}{q}) \sum_a  \exp(- \frac{\dott{\tilde k_a}{\tilde k_a}}{4\l}) 
 g_a (k_i)\, e^{-i k_a \th q} 
\,, 
\end{align}
where 
\begin{align}
\tilde k_\mu &:= \bG_{\mu\nu} \tilde k^\nu := \bG_{\mu\nu} \bar\theta^{\nu\rho} k_\rho \,, \nn\\
\tilde k \cdot \tilde k &= \bar G_{\mu\nu} \thb^{\nu\rho} k_\rho \thb^{\mu\rho'} k_{\rho'}
=  \bLNC^{-4} k_\rho k_{\rho'} g^{\rho\rho'} =: \bLNC^{-4}\,k^2 
\,. 
\end{align}
Basically, $p$ gets replaced by the saddle point. 
There is no UV divergence in the subsequent Schwinger integral at $\l=0$ (see below), 
due to the 4 external $\Theta^{ab}$ vertices. 
The terms $e^{-\frac 1\l \tilde k^2}$ do not change this conclusion provided we keep this  
closed form; an expansion in $\frac 1\l$ would lead to fake UV divergences.
After this Gaussian integration, all sine factors such as $\Sin{k_3\th(k_1+k_2+p)}$
become $\cO(k^2)$, since $p$ gets replaced by a linear combination of the $k_i$.
This takes care of the IR divergence from the propagators
(which arise after integration over the Schwinger parameters), and
there will be no IR divergence as discussed above.
The same argument applies to all higher-order terms $\cO(\Theta^n)$ in the one-loop effective action. 

Carrying out this Gaussian integral explicitly leads to somewhat lengthy expressions.
Introducing the abbreviations 
\begin{align}
 A_2(k) &= \frac 1\l\Big(\bar\a_1 \dott{k_1}{k_1}+\bar\a_2\dott{(k_1+k_2)}{(k_1+k_2)} +\bar\a_3 \dott{k_4}{k_4}\Big) 
- \dott{q(k)}{q(k)} \,, \nn\\
q(k)& = \frac 1\l(\bar\a_1 k_1 + \bar\a_2(k_1+k_2) - \bar \a_3 k_4) \,, \qquad\qquad 
\l=\sum_{i=1}^4 \bar \a_i 
\,, \label{A-def}
\end{align}
where the subscript of $A_2$ emphasizes that it is a quadratic polynomial in the $k_i$, 
we find the following one-loop results after the Gauss integration over $p$ (cf. \eqnref{O4-trace-1}):
\begin{align}
&\int\!\!d^4p\int\limits_0^\infty\!\!d\bar\a_{1-4} \sin\! \left(\frac{k_1 \th  p}{2}\right) \sin\!
   \left(\frac{k_2 \th (p+k_1)}{2} \right) \sin\! \left(\frac{k_3 \th (p+k_1+k_2)}{2}
   \right) \sin\! \left(\frac{p \th (k_1+k_2+k_3)}{2} \right) \nn\\
&\quad \times \exp \left[ - \bar\a_1\dott{(p+k_1)}{(p+k_1)}
   -\bar\a_2 \dott{(p+k_1+k_2)}{(p+k_1+k_2)} \right] \nn\\
&\quad \times \exp \left[-\bar\a_3
   \dott{(p+k_1+k_2+k_3)}{(p+k_1+k_2+k_3)}-\bar\a_4 \dott{p}{p}\right] \nn\\
&= \inv{2^4} \int\limits_0^\infty\!\!d\bar\a_{1-4}\, e^{-\l A_2(k)} \sum\limits_{\zeta_i=\pm1}\frac{\pi ^2 \zeta _1 \zeta _2 \zeta _3 \zeta _4}{\l^2 \sqrt{|g|}} 
   \exp\bigg[-\frac{1}{8 \l}\bigg(\left(1- \zeta _3 \zeta _4\right)
   \dott{\tilde{k}_3}{\tilde{k}_3} \nn\\
&\quad\quad + \left(1-\zeta _1 \zeta_4\right) \dott{\tilde{k}_1}{\tilde{k}_1}
    +\left(1-\zeta _2 \zeta_4\right) \dott{\tilde{k}_2}{\tilde{k}_2}
    +\left(\zeta _1 \zeta_2-\left(\zeta _1+\zeta _2\right) \zeta _4+1\right)\dott{\tilde{k}_1}{\tilde{k}_2} \nn\\
&\quad\quad  +\left(\zeta _1 \zeta_3-\left(\zeta _1+\zeta _3\right) \zeta _4+1\right)
   \dott{\tilde{k}_1}{\tilde{k}_3}+\left(\zeta _2 \zeta_3-\left(\zeta _2+\zeta _3\right) \zeta _4+1\right)
   \dott{\tilde{k}_2}{\tilde{k}_3} \bigg)\bigg] \nn\\
&\quad \exp\bigg[-\frac{i}{2 \l}
   \bigg((k_1 \th{k}_2) \left(\left(\bar\a _2+\bar\a _3\right) \zeta_1
   +\bar\a _4 \zeta _2+\bar\a _1 \zeta _4\right) 
   + (k_2 \th{k}_3) \left(\bar\a _3 \zeta _2+\left(\bar\a_1
   +\bar\a _4\right) \zeta _3+\bar\a _2 \zeta _4\right) \nn\\
&\quad\qquad  
   + (k_1 \th{k}_3) \left(\bar\a _3 \zeta _1+\bar\a _4  \zeta_3+\left(\bar\a _1\!+\!\bar\a _2\right) \zeta _4\right)
  \bigg)\bigg] .
\end{align}
As a consistency check, we note that $A_2 \geq 0$ in the Euclidean case.
To see this, let $\tilde \a_i = \bar \a_i/\l$, so that
\begin{align}
& \(\tilde\a_1 k_1\cdot k_1+\tilde\a_2\Dott{(k_1\!+\!k_2)} +\tilde\a_3 \Dott{k_4}\) 
 \geq (\tilde\a_1 \!+\!\tilde\a_2\!+\!\tilde\a_3) \Big(\frac{\tilde\a_1 k_1+\tilde\a_2(k_1\!+\!k_2) +\tilde\a_3 k_4}
{\tilde\a_1 +\tilde\a_2+\tilde\a_3 }\Big)^{\!2}  \nn\\
& \geq  \(\tilde\a_1 k_1 +\tilde\a_2(k_1+k_2) +\tilde\a_3 k_4\)^2 = \Dott{q(k)}
\,, 
\end{align}
since $\tilde\a_1 +\tilde\a_2+\tilde\a_3  \leq 1$ and using the convexity of the Euclidean norm square.
This implies $A_2 \geq 0$.

In order to carry out at least one of the Schwinger parameter integrals after the Gauss integral, we employ the substitutions
\begin{align}
\bar \a_1 &= \l \xi_1 \xi_2\xi_3 \,, &
\bar \a_2 &= \l (1-\xi_1)\xi_2 \xi_3 \,, \nn\\
\bar \a_3 &= \l(1-\xi_2)\xi_3 \,, & 
\bar \a_4 &= \l(1-\xi_3) \,, \nn\\
\prod d\bar \a_i &= \l^3\xi_2 \xi_3^2 d\l\prod d\xi_i \,, &
 \lambda &= \bar \a_1 + \bar\a_2 +\bar\a_3 + \bar \a_4 \,, 
\label{eq:substitutions}
\end{align}
using
\begin{align}
(\bar\a_1 k_1 + \bar\a_2(k_1+k_2) - \bar \a_3 k_4)
&= \l\xi_3 \(\xi_1 \xi_2k_1 + (1-\xi_1)\xi_2 (k_1+k_2) - (1-\xi_2)k_4\) \nn\\
&=  \l\xi_3 \(\xi_2 k_1 + (1-\xi_1)\xi_2 k_2 - (1-\xi_2) k_4\)  
\,. 
\end{align}
We can now evaluate the $\l$ integral, which gives the following 
higher-derivative and non-local effective action
{\allowdisplaybreaks
\begin{align}
&\varGamma= \tr\big( \tinv{2}\Sigma^{\psi}_{A} \ldots \Sigma^{\psi}_{D} -\Sigma^{Y}_{A} \ldots \Sigma^{Y}_{D}\big)\!
\int\!\!\frac{d^4k_1d^4k_2d^4k_3}{2^{17}\pi^{8}\sqrt{|g|}}\Th^A(-k_1-k_2-k_3)\Th^B(k_3)\Th^C(k_2)\Th^D(k_1)\nn\\
&\times\!\Bigg\{\inv{6}\Bigg(\!\left(8\g_E\!-\!6\right) {\cK_4(k)} 
+ \sum\limits_{i=1}^{4}\! \left(\Dott{\tilde{k_i}}\right)^{\!2} \ln\!\left(\tfrac{\tilde{k_i}\cdot\tilde{k_i}}{4}\right)
- \left(\Dott{(\tilde{k}_1\!+\!\tilde{k}_2)}\right)^{\!2} \ln\!
   \Big(\!\tfrac{ \left(\tilde{k}_1\!+\!\tilde{k}_2\right)\cdot
   \left(\tilde{k}_1+\tilde{k}_2\right)}{4}\Big)\nn\\*
&\; - \left(\Dott{(\tilde{k}_1\!+\!\tilde{k}_3)}\right)^{\!2} \ln\!
   \Big(\!\tfrac{ \left(\tilde{k}_1+\tilde{k}_3\right)\cdot
   \left(\tilde{k}_1+\tilde{k}_3\right)}{4}\Big) 
- \left(\Dott{(\tilde{k}_2\!+\!\tilde{k}_3)}\right)^{\!2} \ln\!
   \Big(\!\tfrac{ \left(\tilde{k}_2+\tilde{k}_3\right)\cdot
   \left(\tilde{k}_2+\tilde{k}_3\right)}{4}\Big)\!\Bigg)
\nn\\
&\; +\int\limits_0^1\!d\xi_{1-3}{\xi _2 \xi_3^2}\Bigg(4 {\cK_4(k)} \ln\left({A_2(k)} \right)  
- 8\,\frac{B_6(k)}{{A_2(k)} } 
\nn\\
&\; +\frac{16 \xi _2\xi_3^2}{{A_2(k)}^2} \left(k_1 \tilde{k}_3-(k_2 \tilde{k}_3) \left(\xi _1-1\right)+(k_1\tilde{k}_2) 
   \xi _1\right) \left((k_1 \tilde{k}_3) \left(\xi _2-1\right)+(k_1 \tilde{k}_2)
   \left(\xi _1 \xi _2-1\right)\right) 
\nn\\*&\;\quad\times\! 
\left((k_1 \tilde{k}_2) \left(\xi _3-1\right)+(k_2 \tilde{k}_3)
   \left(\xi _2-1\right) \xi _3\right)\! \left((k_2 \tilde{k}_3) \left(\left(\xi _1 \xi _2-1\right) \xi
   _3+1\right)-(k_1 \tilde{k}_3) \left(\xi_3-1\right)\right) \!\!
\Bigg)\!\Bigg\} \nn\\
&+\cO(k^6) \, .
\label{lambda-int-result}
\end{align}
Here $A_2(k),B_6(k), \cK_4(k)$ are polynomials in $k_i$ with degree indicated by the subscript. 
They are given by:
}
\begin{align}
\cK_4(k)  &= - \(\dott{\tilde{k}_1}{\tilde{k}_2} \,\dott{\tilde{k}_3}{\tilde{k}_4}
 + \dott{\tilde{k}_1}{\tilde{k}_3}\, \dott{\tilde{k}_2}{\tilde{k}_4} 
+ \dott{\tilde{k}_1}{\tilde{k}_4}\, \dott{\tilde{k}_2}{\tilde{k}_3} \) 
= \frac 14 \Bigg( \sum_{i=1}^4 \tilde k_i^4 - \sum_{i< j \in\{1,2,3\}} (\tilde k_i+\tilde k_j)^4   \Bigg)
, \label{K-id}
\end{align}
while $A_2$ was already defined in \eqref{A-def} which after the substitutions of \eqnref{eq:substitutions} becomes
\begin{align}
A_2(k)&=-\xi _3 \left(\xi _2-1\right) \left(\left(\xi _2-1\right)
   \xi _3+1\right) \Dott{k_4} -\xi _1\xi _2\xi_3 \Dott{k_2}
   +\xi _2^2 \xi _3^2 \Dott{k_1}  \nn\\
&\quad  -2\xi _2 \xi_3^2\big( \left(\xi_2-1\right) \dott{k_1}{k_4}
   + \left(\xi_1-1\right)^2 \xi _2 \Dott{k_2}-2 \left(\xi _1-1\right)\left(\xi _2-1\right)
   \dott{k_2}{k_4}\big) \nn\\ 
&\quad  -2 \xi _2\xi_3\big( \left(\xi_1 + \xi _2 \xi _3 - \xi _2 \xi _3 \xi _1\right) \dott{k_1}{k_2}
   - \Dott{(k_1+k_2)}\big)
\,. 
\end{align}
The remaining trace over the $\Sigma$ in \eqref{lambda-int-result} is computed \appref{app:invariants} resulting in \eqref{four-sigma-trace}. 
Finally, the full expression for $B_6(k)$ is given 
in\footnote{We have not been able to cast these terms into a more transparent form.} \appref{app:suppl}.

\paragraph{Conformal rescaling.}

It is interesting to check the behaviour under conformal rescaling. This is not evident here because
the background $\theta^{\mu\nu}$ breaks the conformal invariance of the $\cN=4$ model. 
Despite their appearance, the $\log$-terms have the correct scaling behaviour  
provided the background is also rescaled appropriately, since
\begin{align}
& 4 {\cK_4} \ln  \left({A_2} \right) + \sum\limits_{i=1}^{4} \left(\Dott{\tilde{k_i}}\right)^2
        \ln\left(\tfrac{\tilde{k_i}\cdot\tilde{k_i}}{4}\right)
 - \sum\limits_{i<j<4}\left(\Dott{(\tilde{k}_i+\tilde{k}_j)}\right)^2 \ln\!
   \Big(\!\tfrac{ \left(\tilde{k}_i+\tilde{k}_2\right)\cdot
   \left(\tilde{k}_1+\tilde{k}_2\right)}{4}\Big) \nn\\
 &= 4 {\cK_4} \ln \! \big(\tfrac{A_2}{\L^2} \big) 
 + \sum\limits_{i=1}^{4} (\Dott{\tilde{k_i}})^{2} \ln\!\big(\L^2\tfrac{\tilde{k_i}\cdot\tilde{k_i}}{4}\big)
 - \sum\limits_{i<j<4}\!\!\left(\Dott{(\tilde{k}_i\!+\!\tilde{k}_j)}\right)^{\!2} \ln\!
   \Big(\!\L^2\tfrac{ \left(\tilde{k}_i+\tilde{k}_2\right)\cdot
   \left(\tilde{k}_1+\tilde{k}_2\right)}{4}\Big) \nn
\end{align}
for any scale $\L$ (in particular $\L=\bLNC$), using the identity \eq{K-id}. Therefore 
these terms are manifestly
invariant under conformal rescaling $k_i \to \a k_i,\, \Theta^A(k) \to \a^{-2}\Theta^A(k)$, 
and also completely symmetric in the external momenta $k_i, \, i=1,2,3,4$.
In fact, this holds true for the complete one-loop effective action due to \eq{four-sigma-trace}. 
Note also that these terms depend only on
the embedding metric $g$, while $A_2$ depends on the effective metric $G$. 
The remaining terms in \eq{lambda-int-result} contain explicit $k_i\theta^{\mu\nu} k_j$ and therefore  violate 
Lorentz invariance manifestly. 
(Lorentz invariance is restored if the $\theta$ are also transformed.) 
Using momentum conservation, all terms of the form $k_i \theta k_i$ with $i,j = 1,\ldots,4$ can be expressed in terms of the 
following basis of anti-symmetric bilinears: 
\be
(k_1+k_2)\theta(k_1+k_3)\,,\qquad (k_1+k_2)\theta(k_2+k_3)\,,\qquad (k_1+k_3)\theta(k_2+k_3) \nn
\,. 
\ee

\section{Non-Abelian sector}
\label{sec:nonabel}
\subsection{General discussion}
\label{sec:nonAB-general}

So far we have studied the quantum effective action for the $U(1)$ sector. 
Now  consider the non-Abelian case, which arises for backgrounds corresponding to 
$N$ copies of coinciding branes with fluctuations,
\begin{align}
X^a &= \bar X^a \, \one_N + \cA^a \qquad \in \algA \otimes \msu(N) \,,  \nn\\
\Theta^{ab} &= \bar \Theta^{ab}\one_N  + \cF^{ab}_\a \l^\a 
\,. \label{nonabel-background}
\end{align}
Here $\l^\a$ is a basis of $\msu(N)$, and 
$\bar X^a \sim \bar x^a: \,\cM^4 \hookrightarrow \R^{10}$ describes an (almost-flat) {\nc} brane $\cM^4$ as before.
Then the fluctuations $\cA^a(x) = \cA^a_\a \l^\a \in \algA \otimes \msu(N)$ describe 
non-Abelian gauge fields\footnote{Recall that any possible $\mU(1)$ fluctuations should be ascribed to the geometry.} and
scalar fields on $\cM^4$. 

If the non-Abelian fields $\cA^a$ have no vacuum expectation value (VEV), 
i.e. if there is no $\msu(N)$ symmetry breaking and $\cN=4$ is preserved,
then the non-Abelian loop modes would give the same contribution to the effective $U(1)$ action as
considered above, starting at $\cO(k^4\Theta^4)$.
However, we now want to consider backgrounds with non-vanishing $\cF^{ab}_\a$.
In that case, there are charged (non-Abelian) fields w.r.t. $\cF$ 
which contribute to the loop integral. This 
will lead to a non-trivial effective action for the $\mSU(N)$ sector
similar to the case of ordinary $\cN=4$ SYM. Furthermore, the associated breaking of $\cN=4$ SUSY 
also leads to an induced  action 
for the $U(1)$ sector at zero momentum, which governs the geometry of $\cM^4$.
Our aim is to extract both effects from 
the one-loop effective action \eq{Gamma-IKKT}, \eq{Gamma-IKKT-4}.

To understand the difference to the Abelian sector more explicitly, note that 
a typical vertex term in the loop integral now looks like 
\begin{align} 
[\cF^{ab}_\a(k_1) e^{i k_1 X}\l^\a,f_\b(k_2) e^{i k_2 X}\l^\b] 
&= -i\Sin{k_1\th k_2} \cF_\a^{ab}(k_1) f_\b(k_2)e^{i (k_1+k_2) X}  \{\l^\a,\l^\b\} \nn\\
& \qquad +  \Cos{k_1\th k_2}\cF_\a^{ab}(k_1) f_\b(k_2) e^{i (k_1+k_2) X} [\l^\a,\l^\b]  \nn\\
&\stackrel{k_i \to 0}{\rightarrow} 
\cF_\a^{ab}(k_1) f_\b(k_2)e^{i (k_1+k_2) X}  [\l^\a,\l^\b]
\,. 
\end{align}
Thus in the low-energy limit $k_i \to 0$, we recover precisely 
the same $\msu(N)$-valued contributions as in the commutative case, while NC effects are sub-leading. 
Since there are no UV divergences in the 
$\cN=4$ model which might induce strong UV/IR mixing, we expect that 
the effective action for the non-Abelian components reduces to that of the commutative 
$\cN=4$  $SU(N)$ gauge theory on a general background 
$\cM^4$, even if that background breaks SUSY. Moreover, the limit $\theta \to 0$ should be smooth
\emph{for the $\mSU(N)$ sector}. 
We emphasize that this statement only applies to the $\cN=4$ resp. the IKKT model.

With this in mind, it makes sense to focus on the limit where $\cF$ is essentially constant along $\cM^4$.
In particular, we can assume that the background satisfies
\be
[\Box,[\cF^{ab},.]] = 0 
\,, \label{box-F-comm}
\ee
which is compatible with the  $SO(10)$ symmetry.
Note that $\Box$ contains also contributions from $\msu(N)$ commutators, nevertheless
this is satisfied for a large class of non-trivial backgrounds, cf. \eq{fuzzy-compatibility}. 
Here the non-Abelian VEV $\cF^{ab}_\a$ is not necessarily tangential to the brane $\cM^4$,
i.e. we admit VEVs both for the non-Abelian field fluxes $F_{\mu\nu}$
as well as $\cF^{\mu i} \sim \bar\theta^{\mu\nu}\del_\nu \Phi^i$.
Then $e^{-\a(\Box + \Sigma_{ab}[\Theta^{ab},.])} = e^{-\a\Box} e^{-\a\Sigma_{ab}[\cF^{ab},.]}$,
dropping $[\bar\Theta^{ab},.]$ which vanishes in the $k_i \to 0$ limit.

To understand the significance of \eq{box-F-comm}, note that both 
 $\Box$ and $[\cF,.]$ are operators on the wave functions 
$\algA \otimes \msu(N)$ which can be simultaneously 
diagonalized. They moreover commute with 
$\Sigma_{ab}$ which acts on the appropriate (spinor, vector or trivial) $\mso(10)$
representation $V$.
Thus the 1-loop low-energy effective action can be written as
\begin{align}
\vG[X]\! 
&:= - \frac 12 \int_0^\infty \frac {d\a}{\a} \Tr\, e^{-\a\Box}\,
    \tr_V\,\Big( e^{-\a\Sigma^{(Y)}_{ab}[\cF^{ab},.]}
  - \frac 12 e^{-\a\Sigma^{(\psi)}_{ab}[\cF^{ab},.] }- 2 \Big) \nn\\
&\,= - \frac 12 \int_0^\infty \frac {d\a}{\a} \Tr \,e^{-\a\Box}\, \chi(-\a[\cF,.])
\,, \label{finite-N-IKKT-1}
\end{align}
which is manifestly $SO(10)$ invariant.
Here, `$\Tr$' is the trace of operators
on $\algA \otimes \Mat(N,\C)$, and 
\begin{align}
\chi_V(\a[\cF,.]) &= \tr_V \big(e^{\a\Sigma^{(V)}_{ab}[\cF^{ab},.]}\big) \,, \nn\\
\chi(\a[\cF,.]) &= \big(\chi_{10}(\a[\cF,.]) - \frac 12 \chi_{16}(\a[\cF,.]) - 2 \big)
= \cO(\a^4) 
\,, \label{chi-def}
\end{align}
is the character of the $SO(10)$ representation $V$, identifying the anti-symmetric matrix 
$[\cF^{ab},.]$ with an (operator-valued) element of $\mso(10)$.

This compact formula is useful in various contexts. To clarify its meaning, 
consider first the semi-classical limit. Then 
the matrix Laplace operator  separates into the tangential covariant Laplacian $\Box_\cM$ on $\cM^4$  
and transversal operator as follows:
\begin{align}
\Box\,\, &\sim\,\, \Box_\cM \, + \, \Box_\perp,  \nn\\
\Box_{\cM} \,\,&=\,\, -\L^{-4}_{\rm NC}\,\nabla^\mu\nabla_\mu \,\,  \sim  \,\, 
[\bar X^\mu + \cA^\mu,[\bar X_\mu + \cA_\mu,.]] \,,   \nn\\ 
\Box_\perp\,\, &= \,\,[\Phi^i,[\Phi_i,.]] 
\,, 
\end{align}
where $\cA^i \equiv \Phi^i$ are the $\msu(N)$-valued fluctuations of $X^a$ perpendicular to $\cM^4$. 
Now recall the commutative\footnote{There is a subtlety: in general, the NC scale 
$\LNC = \LNC(x)$ need not be constant, which would lead to corrections in \eq{SdW-classical}. 
We assume here that $\LNC$ is nearly constant so that 
these corrections can be neglected.} heat-kernel expansion for $\Box_\cM$, 
which for our conventions gives \cite{Gilkey:1995mj,Vassilevich:2003xt}
\be
\Tr (e^{-\a\Box_{\cM}}\, f) \,\, \sim \,\, \frac {\LNC^8}{(4\pi)^2} \frac{1}{\a^2}  
\int_{\cM}d^4x \sqrt{g} \Big(1\, + \cO(\a R, \a F) \Big) f(x)\,
. \label{SdW-classical}
\ee 
As explained in \cite{Blaschke:2010rr}, 
this formula also applies for the trace $\Tr_{\algA}$ over the quantized space of functions $\algA$
in the {\nc} case (recall that $|G|= |g|$ in 4 dimensions),  \emph{provided} 
there is a lower bound\footnote{In particular, there is no good asymptotic expansion as $\a\to 0$ in the NC case.} 
for $\a\geq \a_{\rm min} >0$ corresponding to an effective UV cutoff
\be
\frac{\a_{\rm min}}{\LNC^4} =: \L^{-2} 
\,, 
\ee
such that the momentum scale $k$ of the external field $f$ under the trace \eq{SdW-classical} satisfies
\be
k^2 \L^2 \ll \LNC^4 
\,
\ee
(hence $\a \gg k^2$).
This is justified here because the integral over $\a$ is absolutely convergent at $\a=0$ due to 
maximal supersymmetry, which leads to a factor $\a^4$ in \eq{chi-def} that regularizes the 
$\a$ integral at $\a=0$. 
Then \eq{SdW-classical} holds, because
 the spectral geometry of $\cM^4$ defined by $\Box_\cM$ agrees with the 
classical one corresponding to the effective metric $G$ on $\cM^4$, for the relevant low scales.
Dropping the $\cO( F,R)$ terms from now on, 
the low-energy 1-loop effective action can be written as follows
\begin{align}
\vG[X]\! 
 &\sim - \frac 12\frac {\LNC^8}{(4\pi)^2} \int_0^\infty \frac {d\a}{\a^3} \int_{\cM}d^4x\, \sqrt{g}\, 
 \tr_{\msu(N)} \Big( e^{-\a\Box_\perp} \chi(-\a[\cF,.]) \Big) 
\,. \label{finite-N-IKKT-2}
\end{align}
Although this one-loop action contains no UV divergence, 
there might be standard IR divergences associated with $\a \to \infty$
in the presence of massless gauge fields. To avoid these, we will consider the case of 
spontaneous symmetry breaking where most non-Abelian fields become massive, except possibly some massless $U(1)$ sector.
Then the term $e^{-\a \Box_\perp}$ becomes a mass term for all but the 
unbroken $U(1)$ modes, taking care of the IR divergence.

\subsection{Spontaneous symmetry breaking and massless sector}

Consider a background as above
\be
X^a = \bar X^a\, \one_N + \cA^a  \, ,
\label{eq:general-X-2}
\ee
where $\bar X^a$ defines an (almost-flat) brane $\cM^4$, and  
$\cA^a$ take values in $\msu(N)$ thus breaking the $\msu(N)$ symmetry. 
We denote with $K$ the  unbroken low-energy gauge group resp. Lie algebra defined by the 
transversal scalar fields $\Phi^i$. 
$K$ can be characterized as the kernel of the effective mass operator
\begin{align}
\Box_{\perp} = -[\Phi^i,[\Phi_i,.]] =: \effM^2 
\,, 
\end{align}
which decomposes $\msu(N)$ into eigenspaces with
eigenvalues $m_j^2$,
\be
\msu(N) = K \oplus (\oplus_j V_{m_j^2}) 
\,. \label{sun-decomp}
\ee 
The gauge fields in some non-trivial $V_{m_j^2}$
then acquire a mass due to the Higgs effect, and disappear from the low-energy effective action.
For example, $K = \mmu(1)^{N-1}$ arises if the background fields $\Phi^i$ are proportional to some $\l\in\msu(N)$
with distinct eigenvalues; this case is related to previous work  \cite{Belyaev:2011dg,Banin:2003es,
Buchbinder:2002tb,Buchbinder:2002kn,Lowe:1998vu}
in the commutative $\cN=4$ model.
We wish to determine the low-energy effective action for the massless sector $K$,
along the lines of \cite{Chepelev:1997av}.
However, some of the following considerations are more general.
In particular, we will exhibit the full $SO(10)$ symmetry,
and admit branes with non-trivial geometry.

Assuming $\cF \in K$ it follows that $[\Box,[\cF,.]] =0$, so that
the low-energy effective action starts at $\cO(\cF^4)$ and can be written as 
\begin{align}
\vG_{\textrm{IR}}[X] 
&\sim - \frac{1}2 \frac{\LNC^8}{(4\pi)^2}  \int\limits_0^\infty \frac {d\a}{\a^3} 
   \int_{\cM}\! d^4 x\, \tr_{\msu(N)} \,e^{-\a \effM^2 }\big(
 \chi_{10}(-\a[\cF,.]) - \frac 12 \chi_{16}(-\a[\cF,.]) - 2 \big) \nn\\
&=  - \frac{1}{2}\frac 1{4!} \frac{\LNC^8}{(4\pi)^2}  \int\limits_0^\infty\! d\a\,\a
    \int_{\cM}\! d^4 x\,\tr_{\msu(N)} \,e^{-\a \effM^2 }\Big( (\Sigma^{(Y)}_{ab}[\cF^{ab},.])^4
  - \frac 12 (\Sigma^{(\psi)}_{ab}[\cF^{ab},.] )^4  +\ldots \Big) \nn\\
&= -\frac{1}2 \frac 1{4!}\frac{\LNC^8}{(4\pi)^2} 
   \int_{\cM}\! d^4 x\,\tr_{\msu(N)} \, \inv{\effM^4} \Big( (\Sigma^{(Y)}_{ab}[\cF^{ab},.])^4
  - \frac 12 (\Sigma^{(\psi)}_{ab}[\cF^{ab},.] )^4  +\ldots \Big) 
\label{nonabel-U1}
\end{align}
(see also \appref{app:characters}). 
Decomposing the trace over the $\msu(N)$ according to \eq{sun-decomp},
the massive modes $V_{m_i^2}$ in the loop
now induce a non-trivial contribution to the effective potential for the unbroken modes in $K$,
in contrast to the trace-$U(1)$ sector.  
The reason is that an effective mass is implemented via
\be
e^{-\a\Box_\perp} \equiv e^{-\a \effM^2} 
\ee
in the loop integral. This mass takes care of possible IR problems, while the 
UV divergences are cancelled by the maximal SUSY. 
Note that the massless modes in $K$ do not contribute\footnote{This can be seen from the first line in 
\eq{nonabel-U1}. They do contribute to
higher-derivative terms $O(k^4 \cF^4)$, just like the trace-$U(1)$ considered before.}
in $\tr_{\msu(N)}$
provided $K$ is Abelian, or at least $[\cF,K] = 0$. Then  \eq{nonabel-U1}
is  perfectly well-defined.
If $[\cF,K] \neq 0$, then the $K$ modes in the loop lead to 
a (standard) IR divergence in the non-Abelian sector.
This case is also of interest because it should be related to the non-Abelian DBI action.
We will take care of this problem in \secref{sec:product-spaces}, by considering a geometrical 
background (in the Higgs branch). But first we discuss the Coulomb branch,
where $\cF$ is central in $K$.

\subsection{Coulomb branch}
\label{sec:coulomb-branch}

Consider the case of a $\msu(N)$ gauge theory broken down to 
$K = \msu(N-1) \times \mmu(1)$ through scalar fields $\Phi^i \sim \l$, where 
$\l \sim \diag(1-N,1,\ldots,1)$ is
the generator of the unbroken $\mmu(1)$.
 This can be interpreted as a single ``probe'' brane parallel to a stack of $N-1$
coinciding branes. 
To evaluate the 1-loop effective action for such a background, note that 
\be
 \effM^2 =  (\phi^i \phi_i)\, [\l,[\l,.]]
\label{coulomb-assumptions}
\ee
has only a single non-vanishing eigenvalue $\sim (\phi^i \phi_i)$ with multiplicity 
$2(N-1)$. 
Hence these  $2(N-1)$ modes acquire a mass  $\effM^2 \sim (\phi^i \phi_i)$ in the loop.
Assuming that 
\begin{align}
\cF^{ab} &= \begin{pmatrix}
                             -\theta^{\mu\mu'}\theta^{\nu\nu'} F_{\mu\nu}  & \theta^{\mu\mu'} D_{\mu'}\phi^i \\
                 - \theta^{\nu\nu'} D_{\nu'}\phi^j &  [\phi^i,\phi^j]    \end{pmatrix} \,\l \, ,
\label{theta-explicit-4D}
\end{align}
these massive modes are the only modes
which contribute in \eq{nonabel-U1} since $[\cF,K] = 0$.
We can then evaluate the $\tr_{\msu(N)}$ and obtain 
\begin{align}
\vG_{\textrm{IR}}
&=- (N\!-\!1) \frac{16\LNC^8}{4!(4\pi)^2} \! \int_{\!\cM}\!
  \frac{ d^4 x\,}{(\phi^i \phi_i)^2} \cF^{A} \cF^{B} \cF^{C} \cF^{D}  
\tr\Big(\Sigma^{(Y)}_{A}\Sigma^{(Y)}_{B}\Sigma^{(Y)}_{C}\Sigma^{(Y)}_{D}
 \! - \tfrac 12 \Sigma^{(\psi)}_{A}\Sigma^{(\psi)}_{B}\Sigma^{(\psi)}_{C}\Sigma^{(\psi)}_{D} \Big) \nn\\
&= (N\!-\!1) \frac{\LNC^8}{4!(4\pi)^2} \!\int_{\!\cM}\!
  \frac{d^4 x}{(\phi^i \phi_i)^2}  \cF^{a_1 b_1} \ldots  \cF^{a_4 b_4}\, 6\Big(
  g_{b_1 a_2} g_{b_2 a_1} g_{b_3 a_4} g_{b_4 a_3} \!-4 g_{b_1 a_3} g_{b_3 a_2} g_{b_2 a_4} g_{b_4 a_1} \Big)\nn\\
&=  (N\!-\!1) \frac{\LNC^8}{4(4\pi)^2} \!\int_{\cM}\frac{d^4 x}{(\phi^i \phi_i)^2} \Big( -4  (\cF g \cF g \cF g\cF g)
 +(\cF g \cF g)^2   \Big) \nn\\
&\!\!\!\!\!\!\stackrel{D_\mu \phi^i = 0}{=} \,\, \frac{(N-1)}{4(4\pi)^2} \!\, \LNC^{-8}  
\int_{\cM}\frac{d^4 x}{(\phi^i \phi_i)^2} \Big( -4  (F  G F  G F  G F  G)
 +(F  G F  G)^2   \Big) ,
\end{align}
writing $\cF^{a b}$ for its $\l$ component.
For $D_\mu \phi^i = 0$, we recover 
the well-known $\frac{F^4}{\phi^4}$ term in the low-energy effective action of   
$\cN=4$ SYM for the Coulomb branch \cite{Belyaev:2011dg}.
This term is known to be exact to all orders in perturbation theory, 
as well as non-perturbatively \cite{Dine:1997nq}.
It  vanishes for (anti-)selfdual (ASD) field strength, noting that
\be
4  (F G F G F G F G) - (F G F G)^2 = (F + \star_G F)^2 (F - \star_G F)^2
\,
\label{F+F-indentity}
\ee
where $\star_G$ denotes the 4-dimensional Hodge star with respect to 
the effective metric $G_{\mu\nu}$.
Hence for 4-dimensional NC branes, the induced action for the $U(1)$ brane
is negative (corresponding to an attractive potential), and vanishes identically for (A)SD $F_{\mu\nu}$. 
It is interesting to recall that ASD fluxes also play a preferred role 
for the trace-$U(1)$ sector~\cite{Blaschke:2010qj}.

While the above computation closely follows previous work  \cite{Chepelev:1997av}, our
$SO(10)$ invariant setup also immediately provides the contributions from the scalar fields using
\begin{align}
 \LNC^{8} (\cF g \cF g) &=  F_{\mu\nu} F^{\nu\mu} -2\LNC^{4} D_\mu \phi_i D^\mu \phi^i \,, \nn\\
 \LNC^{16 }(\cF g \cF g \cF g\cF g) & = F_{\mu\nu} F^{\nu\eta} F_{\eta\rho} F^{\rho \mu}  
 -4 \LNC^{4}  D^\mu\phi_i D_\nu\phi^i F^{\nu\eta}  F_{\eta\mu} \nn\\
  &\quad  +2 \LNC^{8} D_\mu\phi_i D_\nu\phi^i  D^\nu\phi_j D^\mu\phi^j
\,, 
\end{align}
where the indices are raised and lowered with  $G^{\mu\nu}$.
Note that  $\phi^i$   
has dimension length here.
This gives 
\begin{align}
\vG_{\rm IR}[X] 
&=  \frac{(N-1)}{4(4\pi)^2} \!\, \LNC^{-8}   \int_{\cM}
  \frac{d^4 x}{(\phi^i \phi_i)^2}  \Big(
 - 4 F_{\mu\nu} F^{\nu\eta} F_{\eta\rho} F^{\rho \mu}  
+ (F_{\mu\nu} F^{\nu\mu} -2\LNC^{4} D_\mu \phi_i D^\mu \phi^i)^2   \nn\\
  &\quad +16 \LNC^{4}  D_\mu\phi_i D_\nu\phi^i F^{\nu\eta} G_{\eta\eta'} F^{\eta'\mu}
  -8 \LNC^{8} D_\mu\phi_i D_\nu\phi^i  D^\nu\phi_j D^\mu\phi^j  \Big) ,
 \label{1-loop-colomb-F4}
\end{align}
whose structure is largely determined by the $SO(10)$ invariance.
Now recall that  the bare matrix model action \eq{bosonic-MM} for non-Abelian 
fields  in the background \eq{theta-explicit-4D} is given by~\cite{Steinacker:2008ya,Steinacker:2010rh} 
\begin{align}
S_{YM}[X] &=-(2\pi)^2\Tr (\co{X^a}{X^b}\co{X_a}{X_b}) \nn\\
&\sim \int_\cM d^4 x\,\LNC^4\, 
 \left(2 D_\mu \phi^i D^\mu\phi_i
 + \LNC^{-4} F_{\mu\nu} F^{\mu\nu} - 2 \eta(x) F \wedge F\right)
\,, \label{eq:1l-eff-YM}
\end{align}
where $\eta(x) = \LNC^{-4}(x) G^{\mu\nu}(x) g_{\mu\nu}(x)$. 
It is remarkable\footnote{This is essentially known in the case of flat branes \cite{Tseytlin:1999dj}.}
that up to this order, after dropping the trace-$U(1)$ terms, the effective action $S_{YM}[X] + \vG_{\rm IR}[X]$ 
 is consistent\footnote{The ``would-be topological terms'' $\eta(x) F \wedge F$ presumably 
correspond to the Chern-Simons terms which should be added to \eq{DBI-action-AdS}.} with the expansion of the 
Dirac-Born-Infeld (DBI) action  
for a D3-brane  in the background of $N-1$ coinciding  D3 branes \cite{Tseytlin:1999dj}
\begin{align}
 S_{\rm DBI} = 
T_3 \int_\cM \!d^4 x\, \frac{|\phi^2|^2}{Q}\left(\sqrt{\left|\det\big(G_{\mu\nu} + \frac{Q}{|\phi^2|^2}D_\mu\phi^i D_\nu\phi_i
+ \frac{Q^{1/2}}{|\phi^2|}\LNC^{-2} F_{\mu\nu}\big)\right|}-\sqrt{\left|\det G\right|}\right) 
\label{DBI-action-AdS}
\end{align}
for $Q=  \frac{(N-1)}{2\pi^2} \LNC^{-4}$ and $T_3 = \LNC^4$. 
In string theory, these constants are given by
 $Q=\frac{(N-1) g_s{\a'}^2}{\pi}$ and $T_3 = \frac 1{2\pi g_s{\a'}^2}$ \cite{Tseytlin:1999dj}.
This is consistent with our results, and allows to identify the scale of non-commutativity
with the string theory parameters as 
\be
g_s {\a'}^2 = \frac{1}{2\pi\LNC^4} 
\label{string-coupling}
\,. 
\ee
Indeed, one can expand \eq{DBI-action-AdS} using 
\begin{align}
\det(G_{\mu\nu} + F_{\mu\nu}) &= \det(G_{\mu\nu}) \Big(1+\inv{2}F_{\m\n}F^{\m\n}-\inv{16}(F_{\m\n}\,\star_G\!F^{\m\n})^2 \Big)
\end{align}
(where indices are raised and lowered with $G_{\mu\nu}$), 
which together with \eq{F+F-indentity} reproduces our results for $S_{YM}[X] + \vG_{\rm IR}[X]$. 

To understand the geometrical meaning of \eq{DBI-action-AdS},
recall that the single D3 brane is modeled by the unbroken $U(1)$ component, 
whose displacement in the transversal $\R^6$ is given by $\phi^i$. Thus
for $\frac{Q}{|\phi^2|^2} \gg 1$ (i.e. for small transversal  distance $|\phi|$ resp. large $N$)
and $G_{\mu\nu} = \eta_{\mu\nu}$, 
the  action \eq{DBI-action-AdS} reduces to the DBI action on a geometry
with effective metric
\begin{align}
ds^2 &= H^{-1/2}(x) dx^\mu dx_\mu + H^{1/2}(x)(d\phi^2 + \phi^2 d\Omega^5)\,, \nn\\
H &= 1+ \frac{Q}{|\phi^2|^2} \;\, \approx \, \frac{Q}{|\phi^2|^2}  \qquad\mbox{for}\,\,\frac{Q}{|\phi^2|^2} \gg 1
\,, 
\end{align}
consistent with IIB supergravity \cite{Buchbinder:2001ui,Maldacena:1997re}.
As is well-known, this reduces to $AdS^5 \times S^5$ in the near-horizon limit.
Our result is in fact more general, since the effective 4-dimensional 
brane metric $G_{\mu\nu}$ is quite generic and not necessarily flat.
Therefore the quantum corrections to the matrix 
model can be interpreted in terms of a modified \emph{bulk} geometry of the embedding space $\R^{10}$.
This illustrates the relation of the matrix model with 
string theory and the gauge-gravity relation. 
The present derivation also applies to the case of generic NC branes with non-trivial embedding geometry. 
For a detailed discussion from the point of view of supergravity in the presence of 
a $B$-field see \cite{Maldacena:1999mh,Alishahiha:1999ci,Berman:2000jw}.

At this point, we recall the general result of emergent gravity~\cite{Steinacker:2010rh} that the effective geometry of a 
(stack of coinciding) branes is given by $G \sim \theta\theta g$,
where $g_{\mu\nu}$ is the pull-back metric on  $\cM^4 \hookrightarrow \R^{10}$. 
However in the above computation, we found that quantum effects 
may modify the effective bulk metric to become $AdS^5 \times S^5$. These seemingly inconsistent results may be 
reconciled as follows: The effective action \eq{DBI-action-AdS} encodes a genuine bulk metric as seen by a test probe 
(i.e. the single D3-brane) at some \emph{distance} $|\phi^2|$ from the $N-1$ coinciding branes. 
This effective bulk metric 
indeed appears to be described by IIB supergravity, in accordance with string theory. 
In the ``emergent gravity'' point of view, it is only the effective metric \emph{on} the brane (or the stack of branes) 
which is of interest, as appropriate in a brane-world picture. To see that we would have to send
the VEV $\phi^i \to 0$, but then the above computation 
 no longer applies as the massive modes become important.
To understand what may happen on several coinciding branes, 
we will discuss in the next section 
the 1-loop effective action in the 
presence of a ``geometrical'' vacuum with $[\phi^i,\phi^j] \neq 0$ or $\cF^{\mu i} \neq 0$, 
such as an extra-dimensional fuzzy space. The effective geometry at tree level is then that of 
$\cM^4 \times \cK \subset \R^{10}$, with induced gravity action for the effective metric
$G_{\mu\nu}$ on $\cM^4$. As an additional bonus, low-energy supersymmetry 
will be generically broken  by the compactification.

In any case, the above computation demonstrates that quantum effects may modify the effective bulk geometry
in the matrix model.
The mechanism is reminiscent of holography based on $\cN=4$ SYM \cite{Maldacena:1997re}, 
but in fact non-commutativity allows a more direct understanding of the branes in the ``holographic'' bulk.
This result underscores the background-independence of the matrix model: There is no 
need to consider different models to describe different background geometries. 
The IKKT model 
is powerful enough to describe non-trivial background geometries not just on the branes, but even for the bulk.
This should clarify the relation of the emergent gravity viewpoint of the matrix model with string theory.

\paragraph{One-loop action and characters.}

For simple backgrounds, the exact one-loop action can be cast into a very simple form,
taking advantage of the formulation in terms of characters. 
For example, consider the case of 
a flat brane $\R^{2n}_\theta$ with constant $\obar\theta^{ab}$ and $\cF^{ab}$. Then
the 1-loop effective action can be written as \eq{finite-N-IKKT-1}
\begin{align}
\Gamma[X] =  -\frac 12 \Tr\int \frac {d\a}\a e^{-\a\Box} \chi(-\a\cF) 
\,. 
\end{align}  
While the kinetic contributions from $e^{-\a\Box}$ in general depend on the geometry, 
the character can be evaluated explicitly in simple cases, 
as explained in \appref{app:characters}. 
If $\cF$ has rank 4, the character is manifestly positive using \eq{chi-rank4},
\begin{align}
 \chi(\a\cF) = \chi_{10} - \tfrac 12 \chi_{16} - 2
= (e^{\frac{\a}2 (f_1-f_2)} - e^{ -\frac{\a}2(f_1-f_2)})^2(e^{\frac{\a}2 (f_1+f_2)} - e^{- \frac{\a}2(f_1+f_2)})^2 .
\nn
\end{align}  
This formula applies in the Abelian case, but is  easily generalized e.g. to the  
Coulomb phase where $\cF^{ab} \sim f^{ab}\l$ for some $\l\in \msu(n)$.
For example, this describes the effective attractive interaction between parallel D3 branes,
which has been shown to be consistent with IIB supergravity \cite{Ishibashi:1996xs}.
In particular $\chi$ is positive definite, 
and vanishes precisely for (anti-)selfdual fluxes $f_1 = \pm f_2$ i.e. in the 
supersymmetric case, cf. \cite{Chatzistavrakidis:2011gs}.
Again the leading term agrees with the DBI action, but there are differences at higher order.

The formulation in terms of characters is also useful in the case of higher rank fluxes. 
For fluxes with higher rank, $\chi(-\a\cF)$ can have either sign depending on the 
specific flux configuration. If two eigenvalues of $\cF$ dominate, then $\chi>0$,
corresponding to a negative potential energy. This should provide a powerful tool to 
select the physically relevant vacua with lowest energy, e.g. in the 
context of brane-world scenarios with intersecting branes in matrix models \cite{Chatzistavrakidis:2011gs}.

\section{Product spaces and SUSY breaking}
\label{sec:product-spaces}

In the previous section, we considered non-Abelian backgrounds
which -- apart from the trace-$U(1)$ sector -- admit a massless unbroken $U(1)$
gauge field. 
We now consider more elaborate backgrounds which 
break the $SU(N)$ gauge symmetry completely, and which can be 
viewed either as product space $\cM^4 \times \cK$, or in terms of a $SU(N)$
gauge theory on $\cM^4$ in the Higgs phase. 
Here  $\cM^4$ is interpreted as (almost-flat, non-compact) space-time, and $\cK$ as compact extra-dimensional
space described by the finite matrix algebra\footnote{The idea to use finite matrix algebras for 
internal spaces is of course not new, cf. \cite{DuboisViolette:1988vq,Aschieri:2006uw}.} 
$\algA_\cK \cong \Mat(N,\C)$. 
We want to understand the 1-loop effective action for such a background,
in particular for the trace-$U(1)$ sector.
The point is that the background will typically break supersymmetry, 
and we expect induced gravity terms.

Thus consider again a background of the form \eq{nonabel-background}
\begin{align}
X^a &= \bar X^a \, \one_N + \cA^a \qquad \in {\algA_{\cM^4}} \otimes \msu(N) \,,  \nn\\
\Theta^{ab} &= \bar \Theta^{ab}\one_N  + \cF^{ab}_\a \l^\a 
\,, \label{nonabel-background-2}
\end{align}
where $\bar X^a$ describes a quantized 4-dimensional manifold $\cM^4$ as in \secref{sec:nonAB-general}. 
However, instead of the Coulomb branch we now assume that the non-Abelian fields 
$\cA^a = \cA^a_\a \l^\a$ resp. $\cF^{ab}$ on $\cM^4$  
describe a quantized compact $2n$-dimensional symplectic (short: fuzzy) space $\cK$. 
In other words, we decompose the full matrix algebra as
\be
\Mat(\infty,\C) \cong {\algA_{\cM^4}} \otimes \mathfrak{u}(N) \cong {\algA_{\cM^4}} \otimes\Mat(N,\C)
\ee
where ${\algA_{\cM^4}}$ is interpreted as algebra of functions on $\cM^4$, and $\Mat(N,\C)$
is interpreted as algebra of functions on $\cK$.
Then the background can be interpreted as higher-dimensional space
\be
\cM = \cM^4 \times \cK \,.
\label{product-BG}
\ee
It is well-known that such $\cK$ may indeed arise from non-Abelian fields on $\cM^4$ 
via the Higgs effect, e.g. the fuzzy sphere $S^2_N$ 
\cite{Madore:1991,Myers:1999ps,Alekseev:1999,Aschieri:2006uw}. 
The curvature of $\cM^4$ is assumed to be small compared with 
the scale of $\cK$, and the embedding of $\cK$
is allowed to vary slowly along $\cM^4$. 
Semi-classically, this means that locally (after a suitable $SO(10)$ rotation) 
$\cM^4 =\R^4_{0123}$ and $\cK \subset\R^6_{456789}$. 
This will be indicated by writing $\bar\Theta^{ab} \equiv \Theta^{\mu\nu}$ and $\cF^{ab} \equiv \cF^{ij}$.  
Then the corresponding Matrix Laplacians
$\Box_{\cM^4} \equiv \Box_4$ and $\Box_\perp \equiv\Box_6$  (almost) commute, and  $[\Theta^{\mu\nu},\cF^{ab}] = 0$
up to sub-leading corrections. 
Thus starting again from \eq{finite-N-IKKT-1}, we can  write
\begin{align}
\Tr(e^{-\a(\Box +[\Theta^{ab},.]\Sigma^{(V)}_{ab}})
 &= \Tr_{{\algA_{\cM^4}}\otimes \algA_\cK\otimes V} 
\Big(e^{-\a(\Box_4  + [\Theta^{\mu\nu},.]\Sigma^{(V)}_{\mu\nu})} e^{-\a(\Box_6 +[\cF^{ij},.]\Sigma^{(V)}_{ij}} \Big) 
\end{align}
where $\Sigma_{ij}$ and $\Sigma_{\mu\nu}$ are generators of  $SO(4)$ resp. $SO(6)$.
The representations of $SO(10)$ decompose accordingly as\footnote{This decomposition may depend on $x\in\cM$.}
\begin{align}
V_{(10)} = V_{(4)} \oplus V_{(6)}\,, \qquad V_{(16,+)} = V_{(2,+)} \otimes V_{(4,+)} \oplus  V_{(2,-)} \otimes V_{(4,-)} 
\,. \label{V-decomp}
\end{align}
This gives
\begin{align}
 \chi_{(10)} = \chi_{(4)} + \chi_{(6)}  
\,, 
\end{align}
while for fermions we have
\begin{align}
\chi_{\rm Dirac(32)} &= \chi_{\rm Dirac(4)} \chi_{\rm Dirac(6)}  \,,  \nn\\
\chi_{(16,+)} &= \chi_{(2,+)} \chi_{(4,+)}  + \chi_{(2,-)} \chi_{(4,-)} 
\,, \label{char-decomp}
\end{align}
cf. \eq{Dirac-character}. The wave-functions can be decomposed accordingly as 
\be
{\algA_{\cM^4}} \otimes \algA_\cK \otimes V_{(10)} = ({\algA_{\cM^4}} \otimes  V_{(4)}) \otimes \algA_\cK 
 \oplus {\algA_{\cM^4}} \otimes (\algA_\cK \otimes V_{(6)}) 
\ee
and similarly for $V_{(16,\pm)}$.
Now let us introduce
\be
\Gamma_{\cM^4,V}(\a) = \Tr_{{\algA_{\cM^4}}\otimes V} \Big(e^{-\a(\Box_4  +[\Theta^{\mu\mu},.]\Sigma^{(V)}_{\mu\nu})} \Big)  
\label{local-model-M}
\ee
where $V$ denotes some representation of $SO(3,1)$.
This is the contribution at scale $\a$ to the effective action 
$\Gamma_{\cM^4,V}(\a) = \int\!\frac{d\a}{\a} \Gamma_{\cM^4,V}(\a)$
of the reduced 4-dimensional spinor/vector/scalar NCFT on $\cM^4$.
Similarly, let
\be
\Gamma_{\cK,V}(\a) = \Tr_{\algA_{\cK}\otimes V} \Big(e^{-\a(\Box_6  +[\cF^{ij},.]\Sigma^{(V)}_{ij})} \Big)
\ee
where $V$ denotes some representation of $SO(6)$, which is the contribution at scale $\a$ to the effective action 
$\Gamma_{\cK,V}(\a) = \int\frac{d\a}{\a} \Gamma_{\cK,V}(\a)$
of the reduced $2n$-dimensional NC model on $\cK$.
Together with \eq{char-decomp} resp. \eq{V-decomp}, it follows that
the low-energy effective action of the full $D=10$ matrix model can be written under the present assumptions as
\begin{align}
\Tr(e^{-\a(\Box +[\Theta^{ab},.]\Sigma^{(10)}_{ab})}) &= \Gamma_{\cM^4,(4)}(\a)\Gamma_{\cK,(1)}(\a) 
  + \Gamma_{\cM^4, 1}(\a)\Gamma_{\cK,(6)}(\a) \,, 
\nn\\
\Tr(e^{-\a(\Box +[\Theta^{ab},.]\Sigma^{(16,+)}_{ab})}) &= \Gamma_{\cM^4,(2,+)}(\a)\Gamma_{\cK,(4,+)}(\a)
                + \Gamma_{\cM^4,(2,-)}(\a)\Gamma_{\cK,(4,-)}(\a) \,, \nn\\
\Tr(e^{-\a\Box} ) &= \Gamma_{\cM^4,(1)}(\a)\Gamma_{\cK,(1)}(\a) 
\,. \label{eff-action-decomp-3}
\end{align}
This decomposition is exact for $\R^4_\theta \times \cK$.
Notice that e.g. in the first line, the $10$ components of the matrix fluctuations 
are  separated into the 4 tangential components
which contribute to $\Gamma_{\cM^4,(4)}\Gamma_{\cK,(1)}$, and 6 transversal components which
contribute to $\Gamma_{\cM^4,(1)}\Gamma_{\cK,(6)}$.
Here $\Gamma_{\cM^4,(4)}\Gamma_{\cK,(1)}$ describes a  NC gauge theory on $\cM$
times a scalar NCFT on $\cK$, and $\Gamma_{\cM, (1)}\Gamma_{\cK,(6)}$ describes a NC scalar theory on 
$\cM$ times a gauge theory on $\cK \subset \R^6$.
Similarly, $\Gamma_{\cM^4,(2,\pm)}\Gamma_{\cK,(4,\pm)}$ is built from spinor models on $\cM^4$ resp. $\cK \subset \R^6$.
Note that the original $U(N)$-valued fields on $\cM^4$ are now interpreted as 
$U(1)$-valued fields on $\cM^4\times \cK$.

This decomposition is physically quite appealing and useful, although the global $SO(10)$ is no longer manifest.
It will provide a better understanding of the
effective 4-dimensional theory at low energy, and allows to see the 
breaking of $\cN=4$ supersymmetry explicitly.

\subsection{Scaling behaviour}

Recall that $\Gamma(\bar\a)$ can be viewed as scale $\L^{-2} \sim \bar\a$ contribution to the one-loop 
effective action $\int_0^\infty \frac{d\bar\a}{\bar\a} \Gamma(\bar\a)$. 
 Assuming a product background as above \eq{product-BG} with a compact fuzzy space $\cK$, there are 2 natural scales
given by the lowest resp. highest Kaluza-Klein (KK) mode on $\cK$.
It is then natural (in the spirit of 
Wilsonian or multi-scale analysis \cite{Rivasseau:1991ub}) to separate the
$\bar\a$ integration into 3 regimes, 
\be
\int_0^\infty d\bar\a = \int_0^{1/\L_1^2} d\bar\a + \int_{1/\L_1^2}^{1/\L_2^2}\ d\bar\a + \int_{1/\L_2^2}^\infty d\bar\a
\,. 
\ee
We choose $\L_1 = N \L_\cK$  to be the UV cutoff scale for $\cK$,
and $\L_2 = \frac 1R \sim \frac 1{N}\L_\cK$ to be the IR cutoff for the compact space $\cK$ with radius $R$.
Then the first integral (``UV regime'')  covers the extreme UV scale above the highest KK state, where 
the modes corresponding to $\cK$ behave as unbroken $SU(N)$ gauge fields on $\cM$ and $\cN=4$ SUSY applies. 
The second integral (``intermediate regime'') covers the scale where $\cK$ behaves as a manifold with NC scale $\L_\cK$, 
describing a $4+2n$-dimensional NC field theory with UV and IR cutoff.
The last integral (``IR regime'') then corresponds to a generically non-SUSY low-energy effective theory 
on $\cM^4$, where all 
non-trivial KK modes are  massive and thus suppressed.

Next, we will describe more explicitly
the $\bar\a$ dependence of the compact contributions in these scaling regimes.
This will allow us to write down effective 4D actions, which contain effective gravitational terms 
as in \cite{Blaschke:2010rr}.

\paragraph{IR regime.} 

In this regime, where $\bar\a^{-1}\leq \L_2^{2}$, it is sufficient to take  into account only the lowest (trivial)
Kaluza-Klein mode on $\cK$. The translational invariance $X^i \to X^i + c^i \one$ of the 
matrix model (which is spontaneously broken by the background geometry but nevertheless
a symmetry of $\Box_6 + \cF^{ij}\Sigma_{ij}$) implies that there are zero modes on $\cK$. 
They correspond to constant wave functions $\sim \one_N$ on $\cK$ and belong to the 
trace-$U(1)$ sector; we will see this explicitly 
in the example of $\cK= S^2_N$ below. Therefore
\begin{align}
\Tr_{\algA_\cK\otimes V} \big(e^{-\a(\Box_6 + \cF^{ij}\Sigma_{ij}} \big) \quad &\stackrel{\a\to\infty}{\to} 
\quad \dim V + N_{0,V}\, e^{-\bar\a m_{0}^2} + \ldots
\,, \label{large-a-contrib-K}
\end{align}
where $m_0$ is the lowest non-trivial eigenvalue of $\Box_6 + \cF^{ij}\Sigma_{ij}$, i.e. the
mass of the lowest non-trivial Kaluza-Klein mode, which is of order
\be
m_0 \sim \frac 1R = \L_2
\,, 
\ee
where $R$ is measured by the effective metric $G$ on $\cK$. 
Note that there are  only finitely many Kaluza-Klein modes
on a fuzzy space $\cK$,  therefore this truncation is justified.

The Kaluza-Klein masses $m_0$ and their multiplicities are in general different 
for the bosonic and fermionic contributions, as we will see in an example below.  
Then SUSY is manifestly broken.
Assuming that the contributions from the two chiral sectors $\Gamma_{\cK,(4,\pm)}(\bar\a)$ coincide,
the 1-loop effective action \eq{eff-action-decomp-3} in the IR regime takes the 
form\footnote{Note that e.g. $\Gamma_{\cK,(4,\pm)}(\bar\a) = 2\Gamma_{\cK,(2)}(\bar\a)$ for the case of a fuzzy sphere,
cf. \eq{Dirac-character}.}
\begin{align}
\Tr(e^{-\a(\Box +[\Theta^{ab},.]\Sigma^{(10)}_{ab})}) &= 
  \Gamma_{\cM,(4)}(\bar\a) (1+N_0 e^{-\bar\a m_{0}^2} + \ldots)  + \Gamma_{\cM, 1}(\bar\a) (6+N_0 e^{-\bar\a m_{0,2}^2} + \ldots) \,, \nn\\
\Tr(e^{-\a(\Box +[\Theta^{ab},.]\Sigma^{(16,+)}_{ab})}) &= 
      \Big(\Gamma_{\cM,(2,+)}(\bar\a) + \Gamma_{\cM,(2,-)}(\bar\a) \Big) (4 + N_0' e^{-\bar\a {m'_0}^{2}} + \ldots)\,, \nn\\
\Tr(e^{-\a\Box}) &= \Gamma_{\cM,(1)}(\bar\a) (1+ N_0'' e^{-\bar\a m_{0}^2} + \ldots) 
\,. \label{eff-action-decomp-3-IR}
\end{align}
Note that the massless contribution combines into the trace-$U(1)$ $\cN=4$ multiplet \eq{Gamma-IKKT}
(which only contributes higher-derivative terms as discussed before which are not of interest here),
but the sub-leading massive contributions are in general no longer supersymmetric. 
Nevertheless, it is useful to write the sub-leading 
contribution in terms of an $\cN=4$ contribution plus some 
(fermionic and/or scalar) correction. Thus we write
the scale $\a$ contribution to the effective action in the IR regime as follows
\begin{align}
\Gamma_{\cM \times \cK}(\bar\a) 
 \,\, = \,\, & (1+N_0 e^{-\bar\a m_0^2} + \ldots)\, \Gamma_{\cM}^{\cN=4}(\bar\a) 
 + (N_0' e^{-\bar\a {m'_0}^{2}} - 4N_0 e^{-\bar\a m_0^2} + \ldots) \Gamma_{\cM}^{\rm fermion}(\bar\a)  \nn\\
&  +  (N_0'' e^{-\bar\a m_{0,3}^2} + \ldots) \Gamma_{\cM}^{\rm scalar}(\bar\a) 
\,. \label{IR-action-expand}
\end{align}
Here $\Gamma_{\cM}^{\cN=4}(\bar\a)$ is the contribution of an Abelian $\cN=4$ matrix model, 
 $\Gamma_{\cM}^{\rm fermion}(\bar\a)$ is the contribution of a fermionic model on  $\cM^4\subset\R^{10}$
and $\Gamma_{\cM}^{\rm scalar}(\bar\a)$ is the contribution of a scalar field model on $\cM^4\subset\R^{10}$.
This organization helps to understand the relevant physics: the first line contains at least 4 derivatives
due to $[\Theta,.]^4$ in the trace-$U(1)$ sector, 
which are of less interest here compared with the
non-vanishing vacuum energy and induced gravity contributions below.
On the other hand, the contributions due to the massive KK modes $\sim e^{-\bar\a m_0^2}$ 
are exponentially suppressed as long as $\L^2 \ll m_0^2$, as shown explicitly in \appref{app:mass-deformed}.
The reason is that these modes are above the UV threshold. In special cases, there may be some massless 
modes as in the example below, but typically SUSY is broken. A gravitational action is then induced
in the intermediate regime, as discussed below.

\paragraph{UV regime.}

In the extreme UV regime, where $\bar\a\L_{UV}^2\ll 1$, we can replace
\begin{align}
\Tr_{\algA_\cK} \big(e^{-\a\Box_6} \big) \quad &\stackrel{\a\to 0}{\to} \quad N^2  \,, \nn\\
\Tr_{\algA_\cK\otimes V} \big(e^{-\a(\Box_6 + \cF^{ij}\Sigma_{ij}} \big) \quad &\stackrel{\a\to 0}{\to} \quad 
 N^2 \dim(V) 
\,, 
\end{align}
because $\a\Box_6 \approx 0$ for all KK modes on $\cK$.
Here $N^2 = \dim(\algA_\cK)$ is finite. 
In particular, this is independent of $\a$. We thus
recover a non-Abelian 4D model with the field content of the full $\cN=4$ SUSY field content
and no symmetry breaking, i.e. 
the original $SO(10)$-invariant $\cN=4$ model. It is UV finite
and induces only higher-derivative terms in the trace-$U(1)$ sector which are not of interest here.
We can therefore simply omit this UV regime.

\paragraph{Intermediate regime.}

For smaller $\bar\a$ corresponding to the second regime, the internal structure of $\cK$ is resolved, and the 
effective action behaves as that of a geometric manifold with $4+2n$ dimensions. 
There are two possible points of view: 
We will first focus on the 4-dimensional point of view keeping track of the KK modes,
and then briefly discuss the higher-dimensional geometrical point of view.

Our main interest is the induced trace-$U(1)$ effective action
due to the finite tower of KK modes in the loop, which become relevant
in the intermediate regime.
Thus consider all terms in \eq{IR-action-expand} involving the higher KK masses $m_i$.
Neglecting first the $m_i$ for simplicity, 
the low-energy action induced e.g. by the fermionic modes in the loop
has been computed in \cite{Blaschke:2010rr} within the matrix model framework:
\begin{align}
\int\limits_{1/\L^{2}}^\infty \frac {d\bar\a}{\bar\a}\,\Gamma_{\cM^4}^{\rm fermion}(\bar\a) 
& =  -\frac 14 \Tr \Bigg(\frac{\L^4\LNC^{-8}}{\sqrt{-\tr J^4 + \frac 12 (\tr J^2)^2 
 +  \L^{-2}\LNC^{4}\cL_{10,\,\textrm{curv}}[X] + \ldots }}\Bigg)  \nn\\
& \sim \intx\, \sqrt{|g(x)|}\Big(\L^4 + \L^2 \cG(x) + \ldots\Big)
\,. \label{full-action-MM}
\end{align}
As expected, one obtains induced vacuum energy and gravity terms\footnote{This was obtained 
in \cite{Blaschke:2010rr}
using a smooth cutoff for the Schwinger parameter rather than a strict cutoff. 
Different implementations of a cutoff are related by some
inessential redefinition of parameters, without altering the conclusions.} 
encoded in the trace-$U(1)$ sector \cite{Blaschke:2010rr}.
Here $J^a_b := i \bar\Theta^{ab} g_{bb'}$, 
and $\cG(x)$ denotes induced gravitational terms which arise from $\cL_{10,\,\textrm{curv}}[X]$, 
including the Einstein-Hilbert term
with additional contributions due to a dilaton and further terms such as $R \theta\theta$.
Terms depending on the extrinsic curvature may also arise.
This formula holds as long as the semi-classical geometric picture is valid,
i.e. 
\be
k^2 \ll \bar\a \LNC^4 \,.
\ee
Here $k$ is the inverse curvature or momentum scale of the background $\cM^4$.

Now let us take into account the $m_i$. Then
each of the (fermionic, say) massive KK modes in the loop induces such a term,
where  $m_i$ acts as an IR cutoff for the loop integral as explained in \appref{app:mass-deformed}.
The UV cutoff also gets modified by the mass term $m_i$,  e.g. $\L^4 \to (\L^4 - m_i^2 \L^2)$; 
however we neglect this modification as we are mainly interested in the qualitative results here.
Thus the contribution due to the KK mode $m_i$ has the structure
\begin{align}
\int\limits_{1/\L_1^{2}}^{1/m_i^2} \frac {d\bar\a}{\bar\a}\,\Gamma_{\cM^4}^{\rm fermion}(\bar\a) 
& \sim \intx\, \sqrt{|g(x)|}\Big((\L_2^4 - m_i^4) + (\L_2^2 - m_i^2) \cG(x) + \ldots\Big)
\,. \label{induced-action-KK-i}
\end{align}
The scalar modes $\Gamma^{\rm scalar}_{\cM}(\bar\a)$ in \eq{IR-action-expand} are also expected to 
induce terms with a similar structure.
As the scale parameter $\bar\a$ approaches the UV scale $\L_1$, the $\cN=4$ SUSY is restored, and these
induced gravitational terms due to the individual KK modes cancel. 
In particular, there is no need to put in any cutoff by hand, and the scaling behaviour discussed above arises 
on physical grounds.
Thus summing up all contributions,
one obtains a \emph{finite} induced vacuum energy and gravity action, as required 
in an induced gravity scenario\footnote{There may be mechanisms other than induced gravity
which play an essential role here, such as the ``harmonic branch'' discussed in \cite{Steinacker:2009mp}.
The main point here is to demonstrate that there are computable and finite induced gravity terms
in the quantum effective action of the matrix model.}. 
The details and even the signs 
of these effective gravity parameters depend on the 
compactification $\cK$, and will not be discussed here.

Finally, consider the geometrical point of view in terms of gauge theory on $\cM = \cM^4 \times \cK$.
Using again the above results on the {\nc}
heat kernel expansion (extrapolated to higher-dimensional cases),
there should be a well-defined expansion for the full heat kernel in $\bar\a$
as long as the IR condition $k^2 \ll \bar\a \LNC^4$, 
i.e. the semi-classical geometric picture is valid. Then the
induced action has the form of a higher-dimensional gravitational action,
\begin{align}
\int\limits_{\L_1^{-2}}^{\L_2^{-2}} \frac {d\bar\a}{\bar\a}\,\Gamma_{\cM}^{\rm fermion}(\bar\a) 
& \sim \int\limits_{\cM^4\times \cK}\!\!\! d^{4+2n}x\, \sqrt{|G(x)|} \Big((\L_2^{4+2n} - \L_1^{4+2n}) 
  + (\L_2^{2+2n}- \L_1^{2+2n}) \tilde\cG(x) + \ldots\!\Big) . \nn
\end{align}
Here $\tilde \cG(x)$ denotes induced $(4+2n)$-dimensional gravitational terms as explained above.
This encodes the non-Abelian sector 
$\algA_\cK \cong \Mat(N,\C)$ in a geometrical manner. From the 4-dimensional point of view, this should be 
reproduced by
computing the induced $SU(N)$ contributions due to the higher KK modes,
providing a cross-check between the geometric and gauge-theory point of 
view. Since a similar check has been done in detail in \cite{Blaschke:2010rr}
in the 4-dimensional case, we do not pursue this any further here.

Let us summarize these observations.
There are higher-derivative contributions to the trace-$U(1)$ sector due to the translational zero modes,
which make up an unbroken $\cN=4$ multiplet. These 
 zero modes are distinct from the  
massive $SU(N)$ modes that break the symmetries and define $\cK$. 
On the other hand, the massive KK modes contribute to $\Gamma_\cK(\bar\a)$ at intermediate scales of the loop integral,
and typically break SUSY. Since these KK modes live on $\cM^4$ and couple 
to the brane metric, they lead to induced gravitational terms in the trace-$U(1)$ sector.
These induced gravitational terms depend to some extent also on the embedding $\cM\subset\R^{10}$, 
since the Dirac operator on the brane is not the standard one. 
Thus one should expect deviations from general relativity, which should arise 
quite generally in similar brane-world scenarios.

We conclude that geometrical backgrounds such as $\cM^4 \times \cK$ do indeed lead to 
induced gravitational 
terms as expected, however their structure is quite non-trivial. Although the details of these terms depend
on the internal space $\cK$, the model is UV finite due to the underlying $\cN=4$ SUSY.
It would be very interesting to study the case of 
more general backgrounds which do not have a product structure. There are indeed solutions of the IKKT model
with the geometry of $\cM^4 \times \cK$ with split non-commutativity,
mixing the compact and non-compact space \cite{Steinacker:2011wb}. 
Although the algebras then no longer factorize, many of the above results are expected to generalize.
Finally, the case of intersecting branes allows to make contact with particle physics \cite{Chatzistavrakidis:2011gs}, 
but poses additional open questions and issues related to gravity which should be studied elsewhere.

\subsection{Example: fuzzy extra dimensions}

To illustrate the above analysis,
let us discuss the example of a fuzzy sphere $\cK \, = \, S^2_N \subset \R^3$ \cite{Madore:1991},
realized by the background $X^i = c_N  \l^i_{(N)}$ where $\l^i_{(N)}$ is the generator of the $N$-dimensional 
irreducible representation of $\msu(2)$.
This is a solution provided\footnote{Lie algebraic solutions of the matrix model without cubic terms do 
exist \cite{Chatzistavrakidis:2011su}, albeit not with the required properties.}
we add the following cubic term $S_{\rm cubic} = \Tr(i C_{ijk} X^i[X^j,X^k])$
to the matrix model action  as in \eq{quad-action-m}, where $C_{ijk}$ is the totally antisymmetric 
symbol corresponding to an embedding of $\mso(3) \subset \R^6$. We use this example merely 
to illustrate the decoupling mechanism discussed above in more detail. There are other
solutions of the ``pure''
IKKT model with compact extra dimensions \cite{Steinacker:2011wb} where analogous considerations 
apply; this will be discussed elsewhere.

First, it is important to note that the trace-$U(1)$ modes 
$X^i \to X^i + c^i\one_N$ correspond to translations of the embedding
of the background $\cM^4 \times \cK$. This is part of the 
algebra $\algA_{\cM^4}$  which describes the geometry 
of $\cM$ rather than $\cK$. This is the sector which leads to UV/IR mixing and enters the 
induced gravity terms on $\cM$.

The algebra $\algA_\cK = \Mat(N,\C)$ of functions on $S^2_N$ decomposes as
\be
\algA_\cK = \Mat(N,\C) \cong (1) \oplus (3) \oplus \ldots \oplus (2N-1)
\ee
under the action of $SU(2)$, and $\Box = [X^i,[X_i,.]] = c_N^2 L^i L_i$ 
is the quadratic Casimir operator, noting that $L_i:= [\l^i_{(N)},.]$ is the angular momentum operator. 
Therefore the contribution from the scalar field is
\begin{align}
\Gamma_{\cK,(1)}(\a) &=  \Tr_{\algA_\cK} \big(e^{-\a\Box_6} \big) 
= \sum_{l=0}^{N-1} (2l+1) e^{- \a \,c_N^2\,l(l+1)}  
 \sim \left\{\begin{array}{ll} 
   N^2, & \a\to 0  \\  
  (1 + 3 e^{-m_0^2\bar\a} + \ldots)  & \a\to \infty  
            \end{array} \right.
\label{gamma-K-ghost}
\end{align}
where $m_0^2 = 2 c_N^2\LNC^4$. 

Similarly for 2-component spinors, 
the space of angular momentum modes is given by
\begin{align}
\algA_\cK\otimes \C^2 &= \Mat(N,\C) \otimes (2) \cong  \Big((1) \oplus (3) \oplus \ldots \oplus(2N-1)\Big) \otimes (2) \nn\\
&\cong  2 \times \Big((2) \oplus (4) \oplus \ldots \oplus (2N-2)\Big) \oplus (2N)  
\,, \label{zeromodes-decomp}
\end{align}
and the relevant Laplacian can be evaluated as
\begin{align}
\Box + \Sigma^{(2)}_{ij}[\Theta_{ij},.] &=  c_N^2 (L_i L^i + 2 S_{i} L^i)
= c_N^2 \Big((\vec L + \vec S)^2 - \frac 34 \Big)
\,. 
\end{align}
This indeed has the expected two zero modes for
``constant'' spinors $(1) \otimes (2)$ due to translational invariance,
and it happens to have 2 additional zero modes where $\vec J = \vec L + \vec S$
has spin $\frac 12$. 
Hence we write 
\begin{align}
\Gamma_{\cK,(2)}(\a)  \sim \left\{\begin{array}{ll} 
           2N^2, & \a\to 0  \\  
  \big(2 + 2 + 8 e^{-m_{0,2}^2\bar\a} + \ldots\big)  & \a\to \infty  
            \end{array} \right.
\,. \label{gamma-K-fermions}
\end{align}
Note that the lowest non-trivial KK mass $m_{0,2}^2 = 3 c_N^2 \LNC^4$, which is different from the 
masses in the bosonic sector. This shows explicitly that SUSY is broken.

Finally consider the bosonic (vector) contribution on $S^2_N \subset \R^3$.
The space of modes is given by
\be
\algA_\cK\otimes (3) = \Mat(N,\C) \otimes (3) 
\,. 
\ee
In order to stabilize the sphere we should actually add a cubic term\footnote{Note that this is consistent 
with translational invariance, but breaks $SO(10)$ to $SO(3)\times SO(7)$.} as in \eq{quad-action-m},
which on a $S^2_N$ background has the same structure as the 
$\Sigma^{(3)}_{ij}[\Theta_{ij},.]$ term. In any case, 
there are three translational zero-modes $(1) \otimes (3)$
which belong to the gravity sector for $\cM^4$. This leads to\footnote{A more detailed discussion of the 
1-loop effective action of such a model has been given in \cite{CastroVillarreal:2004}. Here we only need some basic properties.}
\begin{align}
\Gamma_{\cK,(3)}(\a)  \sim \left\{\begin{array}{ll} 
           3 N^2, & \a\to 0  \\  
  \big(3 + N_0' e^{-m_{0,1}^2 \bar\a} + \ldots\big)  & \a\to \infty  
            \end{array} \right.
\,. \label{gamma-K-bosons}
\end{align}
Similar computations could  be done for other compact NC spaces,
and should be refined for more realistic solutions with extra dimensions 
\cite{Steinacker:2011wb,Chatzistavrakidis:2011gs}.

As a final remark, we point out that
for a large class of NC spaces $\cM$ described by some matrix
algebra $\algA$, the relation 
\be
[\Box,\Sigma_{ij}[\cF^{ij},.]] = 0
\label{fuzzy-compatibility}
\ee
holds. This is obvious for $\R^{2n}_\theta$, but it is also true for  
fuzzy spaces such as $S^2_N$, $\C P^2_N$, etc. Moreover, it is  plausible that this should also hold for 
a large class of physically interesting fluctuations around such spaces, such as on-shell non-Abelian gauge fields
which satisfy $\Box \cF^{ij} = 0$ at least at the linearized level.
 We can then write e.g.
\be
\Tr_{\algA\otimes V} \big(e^{-\a(\Box_4  +[\Theta^{\mu\nu},.]\Sigma^{V}_{\mu\nu})}\big) 
\, = \, \Tr_{\algA} \Big(e^{-\a\Box_4} \tr_V e^{-\a [\Theta^{\mu\nu},.]\Sigma^{V}_{\mu\nu}} \Big)
 =  \Tr_{\algA} \big(e^{-\a\Box_4} \chi_V(\a[\Theta,.]) \big)
\,. 
\ee
This should be useful to determine the low-energy effective action explicitly.

\section{Conclusion}
In this paper, we have continued our study of the one-loop effective action of the IKKT (resp. IIB) 
matrix model, generalizing the analysis of the fermionic induced action given in Ref.~\cite{Blaschke:2010rr}. 
The bosonic part of the action is now included using the background field method, and 
the effective action is elaborated on 4-dimensional {\nc} brane backgrounds. 
We obtain explicit expressions of the leading terms in a momentum expansion for the Abelian 
sector, which are manifestly finite of order $\cO(k^4)$ due to maximal supersymmetry. 
This sector is very different from the commutative $\cN=4$ theory, and governs the geometry of the branes.
In particular, there is no induced gravitational action in the case of unbroken SUSY,
while  scale invariance is broken spontaneously by the background $\theta^{\mu\nu}$ field.

In a second part, we study the effective action for non-Abelian gauge fields, which arise in a background of
$N$ (almost) coinciding NC branes. 
The non-Abelian sector is closely related to standard $\cN=4$ SYM, however
our computations apply also to the case of 4-dimensional branes with general geometry, taking advantage of
recent work on NC branes \cite{Steinacker:2010rh,Blaschke:2010rr}. 
We focus on two cases: the Coulomb branch with an unbroken $U(1)$, 
as well as a background with completely broken gauge group that can be interpreted as a product space
$\cM^4 \times \cK_N$ with fuzzy compact extra dimensions. In the Coulomb branch, 
we show that the effective action coincides with the Dirac-Born-Infeld action for a D3-brane 
in the background of $N-1$ coinciding branes, expanded to leading non-trivial order. 
In particular, the effective bulk metric is seen to be consistent with $AdS^5 \times S^5$
around the stack of branes, 
providing additional evidence for the relation with IIB supergravity. 
The mechanism is also
reminiscent of holography based on $\cN = 4$ SYM, but  non-commutativity allows
a more direct understanding of the branes in the bulk.
Moreover, we obtain a specific 
relation \eq{string-coupling} between the non-commutativity scale and the string coupling.

Finally, we study the effective action on $\cM^4 \times \cK_N$
 in the presence of compact fuzzy extra dimensions,
and give a detailed discussion of the relevant physics at different scales. 
In particular, we demonstrate that supersymmetry can be broken by the extra dimensions $\cK_N$
and their Kaluza-Klein modes, and 
finite gravitational terms are generically induced in the trace--$U(1)$ sector. 
Maximal supersymmetry is restored above a certain scale, ensuring a UV finite effective action.
This supports the picture of emergent gravity on the branes, within a brane-world scenario 
with compactified extra dimensions.

The results of this paper represent another step in the understanding of the IKKT resp. IIB model
at the quantum level. Additional evidence for the relation with IIB supergravity is obtained, 
and the ideas of emergent gravity on NC branes are supported in particular by exhibiting a 
mechanism for breaking $\cN=4$ SUSY. Combined with the recent evidence for a 3+1-dimensional behaviour 
of the IKKT model \cite{Kim:2011cr}, its expected finiteness on 4-dimensional 
brane backgrounds and possible realizations of the standard model  \cite{Chatzistavrakidis:2011gs}, 
there are good prospects to extract real physics 
from this or related matrix models. One interesting extension of this work would be 
to study the one-loop action on the compactified brane solutions found in \cite{Steinacker:2011wb}, which 
could be building blocks towards physically relevant low-energy models.
We hope to report on progress along these lines elsewhere. 
There are many other possible directions of research in this context, which should  
clarify the physical viability of this and related matrix models as a quantum theory of fundamental interactions 
including gravity.

\subsection*{Acknowledgements}

Useful discussions with J. Buchbinder, C-S. Chu, J. Nishimura, 
H. Kawai, Y. Kitazawa,  E. Ivanov, M. Henningson, N. Irges, A. Tseytlin, and A. Wipf are gratefully acknowledged.
D.N. Blaschke is a recipient of an APART fellowship of the Austrian Academy of Sciences. 
The work of H. Steinacker was supported by the Austrian Science Fund (FWF) under contract P21610-N16.

\appendix

\Appendix{Some group theory}
\label{app:group-theory}
\SubAppendix{Invariants}{app:invariants}

We need the traces $\tr \Sigma_A \ldots \Sigma_C$ where $\Sigma_A \equiv \Sigma_{ab}$ denotes the generators of $SO(D)$ 
in the vector or spinor representation of $SO(D)$.
This gives group-theoretical invariants. Obviously $\tr \Sigma_{ab} = 0$.

\paragraph{Quadratic invariants.}

Consider first the quadratic invariant
\begin{align}
\tr \Sigma_A \Sigma_B &= \frac{g_{AB}}{(g_{A'B'} g^{A'B'})}\, \tr\,C^{(2)}
 = \frac{2g_{AB}}{D(D-1)} \tr\, C^{(2)} 
\,. \label{quad-inv}
\end{align}
Here $C^{(2)}$ is the quadratic Casimir, which is easy to compute from $SO(D)$ group theory \cite{Slansky:1981yr}: 
\begin{align}
{C^{(2)}}^{\textrm{spinor}} &=  g^{AB}\Sigma_{A}^{(\psi)}\Sigma_{B}^{(\psi)} = \frac 12 \Sigma_{ab}^{(\psi)}\Sigma_{ab}^{(\psi)} = \frac 18 (D^2 - D)\one \,, \nn\\
{C^{(2)}}^{\textrm{vector}} &=  g^{AB}\Sigma_{A}^{(Y)}\Sigma_{B}^{(Y)} = \frac 12 \Sigma_{ab}^{(Y)}\Sigma_{ab}^{(Y)} = (D-1) \one 
\,, \label{casimir-explicit}
\end{align}
since $A=(a,b);\, a<b$ is a basis of $SO(D)$ and $g^{AB} = \d^{AB}$ is the Killing metric.
Hence
\begin{align}
\frac 12 \tr \Sigma_{ab}^{(\psi)}\Sigma_{ab}^{(\psi)} &= \tr_{\psi} \one\,\frac 18 (D^2 - D) 
 = \frac 18 2^{D/2-1} D(D-1) = 2^{D/2-4} D(D-1) ,  \nn\\
\frac 12 \tr_{Y}  \Sigma_{ab}^{(Y)}\Sigma_{ab}^{(Y)} &= \tr_{Y} \one\, (D-1)  = D(D-1)
\,.
\end{align}
In particular, for $SO(10)$ we get
\begin{align}
\frac 12 \tr \Sigma_{ab}^{(\psi)}\Sigma_{ab}^{(\psi)} &=  180 \,,  &
\frac 12 \tr_{Y}  \Sigma_{ab}^{(Y)}\Sigma_{ab}^{(Y)} &= 90
\,.
\end{align}
Note that the relative factor 2 cancels the explicit $\frac 12$ in front of the fermionic contribution
in \eq{Gamma-schwinger-susy}, so all the $\cO(V^2)$ contributions cancel. 
This shows the special structure of the $\cN=4$ SUSY.

For $SO(6)$, we get
\begin{align}
\frac 12 \tr \Sigma_{ab}^{(\psi)}\Sigma_{ab}^{(\psi)} &= 15 \,, & 
\frac 12 \tr_{Y}  \Sigma_{ab}^{(Y)}\Sigma_{ab}^{(Y)} &= 30
\,. 
\end{align}
Now there is no factor $\frac 12$ in front of the fermionic contribution
in \eq{Gamma-schwinger-susy} because there is no Majorana condition. Therefore
the $\cO(V^2)$ contributions do not cancel, and the model is not one-loop finite but has $\log$ divergences. 
Note that probably $\tr \Sigma\Sigma\Sigma \neq 0$ in that case.

For $SO(8)$, we get
\begin{align}
\frac 12 \tr \Sigma_{ab}^{(\psi)}\Sigma_{ab}^{(\psi)} &=  56 \,, & 
\frac 12 \tr_{Y}  \Sigma_{ab}^{(Y)}\Sigma_{ab}^{(Y)} &= 56
\,,
\end{align}
again no Majorana, hence the $\cO(V^2)$ contributions cancel. 
But recall that the model isn't even supersymmetric and the vacuum energy does not cancel.

\paragraph{Cubic invariants.}

The only cubic invariant tensor in the adjoint is $f_{ABC}$. 
(For $SO(6)$ one might expect a $d_{ABC}$, but this does not seem to arise.) Thus
\be
\tr \Sigma_A \Sigma_B \Sigma_C = \frac i2  \frac{f_{ABC}}{g^{A'B'}g_{A'B'}} \tr\, C^{(2)}
 = i \frac{f_{ABC}}{D(D-1)} \tr\, C^{(2)}
\label{cub-inv}
\ee
since $\tr [\Sigma_A, \Sigma_B] \Sigma_C = i \frac{f_{ABC}}{g^{A'B'}g_{A'B'}} \tr\, C^{(2)} $. 
This gives
\begin{align}
\tr \Sigma_A \Sigma_B \Sigma_C &= i f_{ABC} \frac{D}{D(D-1)}(D-1) =   i f_{ABC}\, , &&\mbox{vector}  \nn\\
\tr \Sigma_A \Sigma_B \Sigma_C &= i f_{ABC} \frac{2^{D/2-1}}{D(D-1)} \frac 18 D(D-1) =  i2^{D/2-4} f_{ABC} \,, &&\mbox{spinor}
\end{align}
which differ by a factor 2 in the $SO(10)$ case,
\begin{align}
\tr \Sigma_A \Sigma_B \Sigma_C &=  i f_{ABC} \,,  \qquad\;  \mbox{vector}  \nn\\
\tr \Sigma_A \Sigma_B \Sigma_C &= 2i f_{ABC} \,, \qquad  \mbox{spinor} .
\end{align}
So even these cubic terms cancel in the IKKT case.
This implies that the model is one-loop finite even on 6D backgrounds, consistent with 
known results on SYM \cite{Green:1982sw,Fradkin:1983jc,Howe:1983jm} which give 
1-loop finiteness in $D<8$.

\paragraph{Quartic invariants.}

The quartic terms no longer coincide: 
\begin{align}
\tr\, C^{(2)} C^{(2)} &= 10 (9)^2 = 810 \,, \qquad\qquad\mbox{vector} \nn\\
\frac 12\tr\, C^{(2)} C^{(2)} &= 8 \(\frac{45}4\)^2 = 1012.5 \,, \qquad\mbox{spinor} 
\end{align}
for $SO(10)$. So these no longer cancel, and we can easily work them out explicitly.
For the vector representation, one obtains
\begin{align}
\Theta^{a_1 b_1} \ldots  \Theta^{a_4 b_4} \tr\big(\Sigma^{(Y)}_{a_1 b_1} \ldots \Sigma^{(Y)}_{a_1 b_1}\big) 
 &= 16 \Theta^{a_1 b_1} \ldots  \Theta^{a_4 b_4}\, g_{b_1 a_2} g_{b_2 a_3} g_{b_3 a_4} g_{b_4 a_1} \nn\\
&= 16 \Theta^{a b} \Theta^{b c}  \Theta^{c d}  \Theta^{d a}
\,. 
\end{align}
To compute the spinorial traces, we use the standard contraction formulas for Dirac spinors
\begin{align}
&\tr\Big(\g_{a_1} \ldots \g_{a_{2n}} \Big) 
 =  \sum_{\rm contractions\, \cC} (-1)^{\cC} g_{a_ia_j} \ldots g_{a_ka_l} \tr\one  \,,  \nn\\
&\tr\Big(\g_{a_1} \ldots \g_{a_{2n}} \gamma\Big) = 0\,, \qquad  2n < 10 \,, \nn\\
&\tr_{\rm Dirac}\Big(\g_{a_1} \ldots \g_{a_{2n}}(\frac{1+\gamma}2)\Big) 
= 16\, \sum_{\rm contractions\, \cC} (-1)^{\cC} g_{a_ia_j} \ldots g_{a_ka_l} , \quad  2n < 10 
\,, 
\end{align}
where $\gamma = i\g_1 \ldots \g_{10}$ in the Euclidean case. 
For 10 insertions, the $\gamma$  gives an imaginary contribution to the effective action
with 10 insertions, which is characteristic for the WZ term.

For the chiral spinor representations of $SO(10)$, this gives
\begin{align}
& \Theta^{a_1 b_1} \ldots  \Theta^{a_4 b_4} \tr\big(\Sigma^{(\psi)}_{a_1 b_1} \ldots \Sigma^{(\psi)}_{a_4 b_4}\big) \nn\\
&= 4 \Theta^{a_1 b_1} \ldots  \Theta^{a_4 b_4}\, \Big(
4 g_{b_1 a_2} g_{b_2 a_3} g_{b_3 a_4} g_{b_4 a_1} 
- 4 g_{b_1 a_2} g_{b_2 a_4} g_{b_4 a_3} g_{b_3 a_1} 
- 4 g_{b_1 a_3} g_{b_3 a_2} g_{b_2 a_4} g_{b_4 a_1} \nn\\
&\quad +  g_{b_1 a_2} g_{b_2 a_1} g_{b_3 a_4} g_{b_4 a_3}
 +  g_{b_1 a_3} g_{b_3 a_1} g_{b_2 a_4} g_{b_4 a_2}
+  g_{b_1 a_4} g_{b_4 a_1} g_{b_2 a_3} g_{b_3 a_2}
 \Big)
\end{align}
using $\frac{\tr \one}{16}\,=1$. The first 3 terms are connected contractions while the last 3 are disconnected 
contractions, which have a symmetry factor 4. 
This gives
\begin{align}
& \Theta^{a_1 b_1} \ldots  \Theta^{a_4 b_4} 
 \tr\Big(\frac 12 \Sigma^{(\psi)}_{a_1 b_1} \ldots \Sigma^{(\psi)}_{a_4 b_4} - \Sigma^{(Y)}_{a_1 b_1} \ldots \Sigma^{(Y)}_{a_4 b_4}\Big) \nn\\
&= 2 \Theta^{a_1 b_1} \ldots  \Theta^{a_4 b_4}\, \Big(-4 g_{b_1 a_2} g_{b_2 a_3} g_{b_3 a_4} g_{b_4 a_1} 
- 4 g_{b_1 a_2} g_{b_2 a_4} g_{b_4 a_3} g_{b_3 a_1} 
- 4 g_{b_1 a_3} g_{b_3 a_2} g_{b_2 a_4} g_{b_4 a_1} \nn\\
&\quad +  g_{b_1 a_2} g_{b_2 a_1} g_{b_3 a_4} g_{b_4 a_3}
 +  g_{b_1 a_3} g_{b_3 a_1} g_{b_2 a_4} g_{b_4 a_2}
+  g_{b_1 a_4} g_{b_4 a_1} g_{b_2 a_3} g_{b_3 a_2} \Big) \nn\\
&= 6 \, \Big(-4 (\Theta g \Theta g \Theta g \Theta g) 
 +  (\Theta g \Theta g)^2 \Big) 
\,, \label{four-sigma-trace}
\end{align} 
 for the IKKT model,
which is not only cyclic but in fact totally symmetric in the $\Theta^{ab}$.

\SubAppendix{Characters}{app:characters}
\noindent
The character for a representation $V$ of some Lie algebra $\mg$ is defined as 
\be
\chi_V(H) = \Tr\, e^{H} 
\,, 
\ee
where $H\in \mg$ (which is often assumed to be in the Cartan sub-algebra and thus identified with a weight).
Characters are very useful objects in group theory, notably because they satisfy
$\chi_{V \otimes W} = \chi_V \chi_W$. In the present context, 
we can interpret the term $\tr e^{\a \Sigma_{ab}[\Theta^{ab},.]}$
as character of $SO(10)$, for a  given matrix $\Theta^{ab}$ (one may also try to interpret $[\Theta^{ab},.]$
as a generator of $\msu(N^2)$ acting on the space of matrices).

For a given flux $\cF^{ab}$ (either constant or at some point in the semi-classical limit),
 we can choose a basis  using a suitable 
$SO(D)$ rotation where $\cF^{ab}$ is block-diagonal:
\begin{align}
\cF^{ab} \sim \begin{pmatrix} 0 & f_1 & & & \\
                       -f_1 & 0  & & &  \\
                   & & \ddots & & \\
                      &  & & 0 & f_5 \\
                      &  & & -f_5 & 0
 \end{pmatrix}
= \begin{pmatrix} f_1 \, i\sigma_2 & & \\
                   & \ddots  & \\
                        & 0 & f_5\,i\sigma_2 \\
 \end{pmatrix}
. 
\end{align}
In this basis, we can then choose a corresponding  fermionic oscillator rep. for the Gamma matrices,
\begin{align}
2\a_i &= \gamma_{2i-1} - i \gamma_{2i}\,, \qquad 2\a_i^+ =  \gamma_{2i+1} + i \gamma_{2i}\,, \qquad
\{\a_i,\a_j^+\} = \d_{ij}\,, \qquad i=1,2,\ldots,5 \,, \nn\\
i\gamma_1\gamma_2 &= -2(\a_1^+ \a_1-\frac 12)\,,  \qquad\qquad
 i\gamma_3\gamma_4 =  -2(\a_2^+\a_2-\frac 12)\,, \,\mbox{etc.}, \nn\\
\Sigma_{12} &= \frac i4 [\gamma_1,\gamma_2] = -\frac 12[\a_1,\a_1^+] = -\frac 12(1-2\a_1^+\a_1) =  \frac 12\chi_{(1)}
= \frac 12 \sigma_3
\,, 
\end{align}
etc. which act on the spin $\frac 12$ irrep.
\begin{align}
\cF^{ab}\Sigma^{ab} &= f_1 \sigma_3 \otimes f_2 \sigma_3 \otimes \ldots \otimes f_5 \sigma_3 \,,  \nn\\
\chi_{(32)}(\a\cF) = \tr_{32}(e^{\a\cF^{ab}\Sigma^{(\psi)}_{ab}}) &= (e^{\a f_1} + e^{-\a f_1}) \ldots (e^{\a f_5} + e^{-\a f_5})
 = \sum_{n_i=\pm 1} e^{\a n_i f_i} 
\,, \label{Dirac-character}
\end{align}
which acts on $\C^{32}$.
The most general state for a Dirac fermion $\Psi$ can  be written as 
\be
\Psi = \sum_{n_i = \pm 1}  \psi_{n_1 \ldots n_5} |n_1 \ldots n_5\rangle \psi_{n_1 \ldots n_5} 
\ee
where the ket denotes spinor states.
It is the sum of both chiral contributions. 
\begin{align}
&\chi_{(16,+)}(\a\cF) = \sum_{n_i=\pm 1,\sum n_i = 5,1,-3}  e^{\a n_i f_i} \nn\\
 &= e^{\a (f_1+f_2+f_3+f_4+f_5)}+  e^{\a (f_1+f_2-f_3-f_4+f_5)} + e^{\a (f_1+f_2-f_3-f_4-f_5)} + e^{\a (f_1+f_2+f_3-f_4-f_5)}  \nn\\
&\quad +  e^{\a (-f_1-f_2+f_3+f_4+f_5)} + e^{\a (-f_1+f_2-f_3+f_4+f_5)} + e^{\a (-f_1+f_2+f_3-f_4+f_5)} + e^{\a (f_1+f_2+f_3+f_4-f_5)} \nn\\
&\quad +  e^{\a (f_1-f_2-f_3+f_4+f_5)} + e^{\a (f_1-f_2+f_3-f_4+f_5)} + e^{\a (f_1-f_2+f_3+f_4-f_5)} \nn\\
&\quad + e^{\a (f_1-f_2-f_3-f_4-f_5)} + e^{\a (-f_1+f_2-f_3-f_4-f_5)} + e^{\a (-f_1-f_2+f_3-f_4-f_5)}\nn\\
&\quad + e^{\a (-f_1-f_2-f_3+f_4-f_5)} + e^{\a (-f_1-f_2-f_3-f_4+f_5)} \,,  \nn\\ 
&\chi_{(16,+)}(\a\cF)  = \sum_{n_i=\pm 1,\sum n_i = 3,-1,-5}  e^{\a n_i f_i} 
\,. \label{Weyl-character}
\end{align}
Notice that $\tr_{16,\pm}(e^{\a\cF^{ab}\Sigma^{(\psi)}_{ab}}) = \frac 12 \tr_{32}(e^{\a\cF^{ab}\Sigma^{(\psi)}_{ab}})$
whenever $\cF$ has rank at most 8, since then both contributions from $e^{\pm\a f_5}$ coincide.

On the other hand, on the vector representation we have 
\begin{align}
\chi_{(10)}(\a\cF) =\tr(e^{\a\cF^{ab}\Sigma^{(Y)}_{ab}}) &= \sum_i (e^{2\a f_i} +  e^{-2\a f_i}) 
\,. \label{Vector-character}
\end{align}
In the case of a rank 4 flux, this gives 
\begin{align}
 \tr_{10}(\a\cF) - \frac 14 \tr_{32}(\a\cF^{ab}) - 2
&= (e^{2\a f_1} + e^{-2\a f_1} +e^{2\a f_2} + e^{-2\a f_2} +6) \nn\\
  &\quad - \frac 14 8 (e^{\a f_1} + e^{-\a f_1})(e^{\a f_2} + e^{-\a f_2})  - 2 \nn\\
&= (e^{\a (f_1-f_2)/2} - e^{-\a (f_1-f_2)/2})^2(e^{\a (f_1+f_2)/2} - e^{-\a (f_1+f_2)/2})^2\nn\\
& \geq 0
\label{chi-rank4}
\end{align}
which is positive definite and vanishes precisely for (A)SD fields. 
This is consistent with previous results.

\vspace*{1.5em}

\Appendix{Mass deformed matrix model}
\label{app:mass-deformed}

Adding a mass term to the IKKT model does not add any potential or gravitational terms to the 
$U(1)$ sector of the effective action,
because the $[\Theta,.]^4$ terms involve at least 4 derivatives.

Consider e.g. the following mass-deformed matrix model where only the fermionic part is massive, 
i.e. the bosonic and ghost parts are massless. 
A computation along the same lines as in \eqnref{Gamma-IKKT} then leads to 
\begin{align}
 \vG[X,m^2] &:= - \frac 12 \Tr \int\limits_0^\infty\! \frac {d\a}{\a} 
    \Big( e^{-\a( \Box  + \Sigma^{(Y)}_{ab}[\Th^{ab},.])}
   - \frac 12 e^{-\a(\bar m^2 + \Box  + \Sigma^{(\psi)}_{ab}[\Th^{ab},.]) }
  - 2 e^{-\a(\Box)}  \Big) \nn\\
&= \vG[X,0]  + \frac 14 \Tr \int\limits_0^\infty\! \frac {d\a}{\a} 
\( e^{-\a(\bar m^2 + \Box  + \Sigma^{(\psi)}_{ab}[\Th^{ab},.]) } - e^{-\a(\Box  + \Sigma^{(\psi)}_{ab}[\Th^{ab},.]) }\)
 \end{align}
where $\vG[X,0]$ is the contribution of the massless $\cN=4$ model.
Thus we need to understand the effect of such a mass term in the integrand.
For simplicity, consider the contribution of a massive scalar field on a flat space
\begin{align}
\Tr \int\limits_{1/\L^2}^\infty\! \frac {d\a}{\a}  e^{-\a( \Box + \bar m^2)} 
 &=  \Intt{p} \int\limits_{1/\L^2}^\infty\! \frac {d\bar\a}{\bar\a}  e^{-\bar\a(p^2  + m^2)}  \nn\\
&= \frac 1{4\LNC^4}  \int\limits_{1/\L^2}^\infty\! \frac {d\bar\a}{\bar\a^3} \, e^{-\bar\a m^2}  
  \approx \, \frac 1{4\LNC^4}  \int\limits_{0}^\infty\! \frac {d\bar\a}{\bar\a^3} \, e^{-\bar\a m^2- \frac 1{\a\L^2}}  
\end{align}
where $\bar m^2 = \LNC^{-4}  m^2$, and 
introducing a UV cutoff $\L$. We will use the latter form of a UV cutoff, which is slightly more convenient.
This can be evaluated using 
\begin{align}
&\int_{0}^{\infty}\!d\a\, \frac{1}{\a^n}\exp\Big(-m^2 \a - \frac 1{\a\L^2}\Big)
 = 2(\L m)^{n-1}\, K_{n-1}\Big(\frac{2m}{\L}\Big)  
\,. 
\end{align}
Now we can use the asymptotic behaviour of the Bessel functions 
\begin{align}
K_n(x) \sim \left\{ \begin{array}{ll}e^{-x}\Big(\sqrt{\frac{\pi}{2x}} + \cO(x^{-3/2})\Big)\,, & x \to\infty\\
          c_n \(\frac 2x\)^n - c_{n-1} \(\frac 2x\)^{n-2}+ \cO(x^n)  \,, & x \to 0        
                    \end{array} \right.
\,. 
\end{align}
Therefore, below the mass threshold $\L^2 < m^2$, the induced action is exponentially suppressed
since the massive KK modes are not excited, while for 
$\L^2 > m^2$ the mass becomes irrelevant and the standard vacuum energy contributions 
(cosmological const., induced Einstein-Hilbert term, etc.) are induced. More precisely, 
$K_2(x) \sim \frac 2{x^2} - \frac 12 + \cO(x^2)$ gives 
\begin{align}
&\int_{0}^{\infty}\!d\a\, \frac{1}{\a^3}\exp\Big(-m^2 \a - \frac 1{\a\L^2}\Big)
  = \L^4 - \L^2 m^2 + \cO(m^4)
\end{align}
so that the cutoff is effectively modified by the mass term as $\L^4\to\L^4-m^2\L^2 + \cO(m^4)$ for $\L^2 \gg m^2$.
Similarly, we find $\L^2 \to \L^2 + \cO(m^2\log\L)$, and 
$\L^6\to\L^6-\frac 12 m^2\L^4 + \cO(m^4 \L^2)$ for higher-order terms in 
the induced action as in Ref.~\cite{Blaschke:2010rr}.

\Appendix{Supplementary results}
\label{app:suppl}
For completeness, we give the explicit expression for the abbreviation $B_6(k)$ introduced in \eqnref{lambda-int-result}:
\begin{align}
&B_6(k)=\nn\\
&\bigg(\! (k_1 \tilde{k}_2)^2 \xi _3 \!\left(\dott{\tilde{k}_1}{\tilde{k}_3} \left(2 \xi _1 \xi
   _2-1\right) (\xi_3\!-\!1)+\left(\xi _1 \xi _2-1\right)
   \left(\Dott{\tilde{k}_3} (\xi_3\!-\!1)+\dott{\tilde{k}_2}{\tilde{k}_3} \left(\xi _1 \xi _2 \xi
   _3+\xi _3-1\right)\right)\!\right)
\nn\\*
& +(k_2 \tilde{k}_3)^2 \xi _3
   \left(\dott{\tilde{k}_1}{\tilde{k}_2} \left(\left(\xi _1-2\right)
   \xi _2+1\right) \left(\xi _1 \xi _2 \xi _3-\xi
   _3+1\right)-\Dott{\tilde{k}_1} \left(\xi _2-1\right)
   \left(\left(\xi _1 \xi _2-1\right) \xi_3+1\right)
\right.\nn\\*&\quad \left.
-\dott{\tilde{k}_1}{\tilde{k}_3} \left(\xi _2-1\right)
   \left(\left(2 \xi _1 \xi _2-\xi _2-1\right) \xi
   _3+1\right)\right)
\nn\\*
& + (k_1 \tilde{k}_3)^2 \xi _3
   \left(\dott{\tilde{k}_1}{\tilde{k}_2} \left(2 \xi _2-1\right)
   \left(\xi _3-1\right)+\left(\xi _2-1\right)
   \left(\Dott{\tilde{k}_2} \left(\xi_3-1\right)+\dott{\tilde{k}_2}{\tilde{k}_3} \left(\xi _2 \xi _3+\xi
   _3-1\right)\right)\right)
\nn\\*
& +(k_1 \tilde{k}_2)(k_2 \tilde{k}_3)
   \!\left(\dott{\tilde{k}_1}{\tilde{k}_3}-\Dott{\tilde{k}_1}
   (\xi_3\!-\!1) \left(\left(\xi _1 \xi _2\!-\!1\right) \xi
   _3\!+\!1\right)-\dott{\tilde{k}_1}{\tilde{k}_2} \left(2 \xi _1 \xi _2
   \xi _3\!-\!1\right) \left(\left(\xi _1 \xi _2\!-\!1\right) \xi_3\!+\!1\right)
\right.\nn\\*&\quad\left.
+\xi _3 \left(\dott{\tilde{k}_2}{(\tilde{k}_2+\tilde{k}_3)}+\Dott{\tilde{k}_3}
   \left(\xi _2-1\right) \left(\xi _1 \xi _2-1\right)
   \xi _3-\dott{\tilde{k}_2}{\tilde{k}_3} \xi _2 \left(\xi _1+2
   \left(\xi _1-1\right) \left(\xi _1 \xi _2-1\right) \xi
   _3\right)\right)
\right.\nn\\*&\quad\left.
+\xi_3\left(-\Dott{\tilde{k}_2} \left(\xi _3 \left(\xi _1
   \xi _2-1\right)^2+\xi _1 \xi _2\right)+\dott{\tilde{k}_1}{\tilde{k}_3} \left(2 \xi _1 \xi _2-\xi _2+2 \left(\xi _1
   \left(\xi _2-2\right) \xi _2+1\right) \xi_3-2\right)\right)\right)
\nn\\*
& +(k_1 \tilde{k}_3)(k_1 \tilde{k}_2)
   \!\left(\Dott{\tilde{k}_1} (\xi_3\!-\!1)^2+\dott{\tilde{k}_1}{\tilde{k}_2} \left(2 \xi _1 \xi
   _2 \xi _3\!-\!1\right)\! (\xi_3\!-\!1)+\dott{\tilde{k}_1}{\tilde{k}_3}
   +\xi _3 \dott{\tilde{k}_1}{\tilde{k}_3} \left(2 \xi _2 (\xi _3\!-\!1)\!-\!1\right)
\right.\nn\\*&\quad\left.
+\xi _3 \left(\!\left(\Dott{\tilde{k}_3}
   (\xi_2\!-\!1) +\Dott{\tilde{k}_2} \left(\xi _1 \xi_2\!-\!1\right)\right) (\xi_3\!-\!1)+\dott{\tilde{k}_2}{\tilde{k}_3}
   \left(-2 \xi _3\!+\!\xi _2 \left(2 \xi _2 \xi _3 \xi _1\!-\!\xi_1\!-\!1\right)+2\right)\!\right)\!\right) 
\nn\\*
& +(k_1 \tilde{k}_3) (k_2 \tilde{k}_3) \xi _3 \left(\Dott{\tilde{k}_1}
   \left(\xi _2-1\right) \left(\xi _3-1\right)+\dott{\tilde{k}_1}{\tilde{k}_2}
   \left(-2 \xi _3+\xi _2 \left(-2 \xi _2 \xi _3 \xi_1+\xi _1+4 \xi _3-4\right)+2\right)
\right.\nn\\*&\quad\left.
 +\left(\xi _2-1\right)
   \left(\Dott{\tilde{k}_3}{\tilde{k}_3} \left(\xi _2-1\right) \xi
   _3+\dott{\tilde{k}_1}{\tilde{k}_3} \left(2 \xi _2 \xi
   _3-1\right)+\dott{\tilde{k}_2}{\tilde{k}_3} \left(-2 \left(\xi
   _1-1\right) \xi _2 \xi _3-1\right)\right)
\right.\nn\\*&\quad\left.
+\left(\xi _2-1\right)\left(\Dott{\tilde{k}_2} \left(-\xi _1 \xi _2 \xi _3+\xi _3-1\right)\right)\right)
\!\bigg)\,. 
\end{align}


\bibliographystyle{../custom1.bst}
\bibliography{../articles.bib,../books.bib}

\end{document}